These are the accepted versions of two book chapters, provided prior to final layout and copyediting. Please refer to the published book for the **final** versions and cite as follows:

# Communication as a driver of change

Prof. Andrea Fronzetti Colladon, Department of Civil, Computer Science and Aeronautical Technologies Engineering, Roma Tre University, Via Vito Volterra 62, 00146 - Rome, Italy. ORCID: 0000-0002-5348-9722

Prof. Francesca Grippa, Northeastern University, 360 Huntington Avenue, Boston, MA, 02115, USA. ORCID: 0000-0003-3537-6326

**Abstract**

Communication plays a pivotal role in driving change, influencing human behavior, and solving problems. This chapter explores the pragmatic power of communication, drawing examples and research to illustrate its impact on people's perceptions and actions. The chapter examines how varied perceptions of reality impact responses and highlights the importance of understanding others' viewpoints to foster effective change. Key topics covered in this chapter include goal setting, dysfunctional solutions, and clouded perceptions. It also illustrates unconventional logics, such as paradoxes and contradictions, which can help break through emotional barriers. Practical tools and exercises are provided within a structured framework for guiding both personal and organizational change, offering insights on managing emotional and rational aspects to unlock potential and drive meaningful transformation.

**Keywords:** change management; communication; problem-solving; goal setting; dysfunctional solutions; unconventional logics.

In ancient times, words were regarded as precious and powerful, carefully chosen by sorcerers to heal illnesses or communicate with the divine. Even today, words retain much of their ancient, almost magical power (S. Freud, 1989). Through words, a company can inspire action in consumers, while an individual can evoke profound joy or deep despair in another. As such, our communication holds the potential to foster or hinder change, drive transformation, and solve problems (Watzlawick, Weakland, et al., 2011). Research has shown that communication has a pragmatic effect (Watzlawick, Bavelas, et al., 2011), meaning it can influence behaviors. In this chapter, we explore how communication can be a driver of change and its potential effects on human behavior.

We do not cover propaganda or misinformation, even though they are hot topics nowadays. Instead, we provide a general overview of the kinds of communication and human interaction that can unlock jammed mechanisms and unleash the talents and potential of individuals.

The first step in understanding how people behave is acknowledging that each of us lives in our own representation of reality. Therefore, we should not assume the existence of a unique truth but rather try to unveil how different people represent what they observe, the stimuli they receive from their inner self, other individuals, and the environment (e.g., a text message, a smile, a pat on their back).

In general, we can discuss the existence of a *perception-reaction system* that varies from person to person. Consequently, diverse reactions arise due to distinct perceptions or interpretations of reality. This way of interpreting what is *true* or *false* can be found in the work of relativists and in the views attributed to Protagoras (Lee, 2005). Accordingly, everyone evaluates things based on how they perceive them and their personal history, expectations, and experiences developed over a lifetime. For example, a storm would scare a child and make a country holder happy after a period of drought.

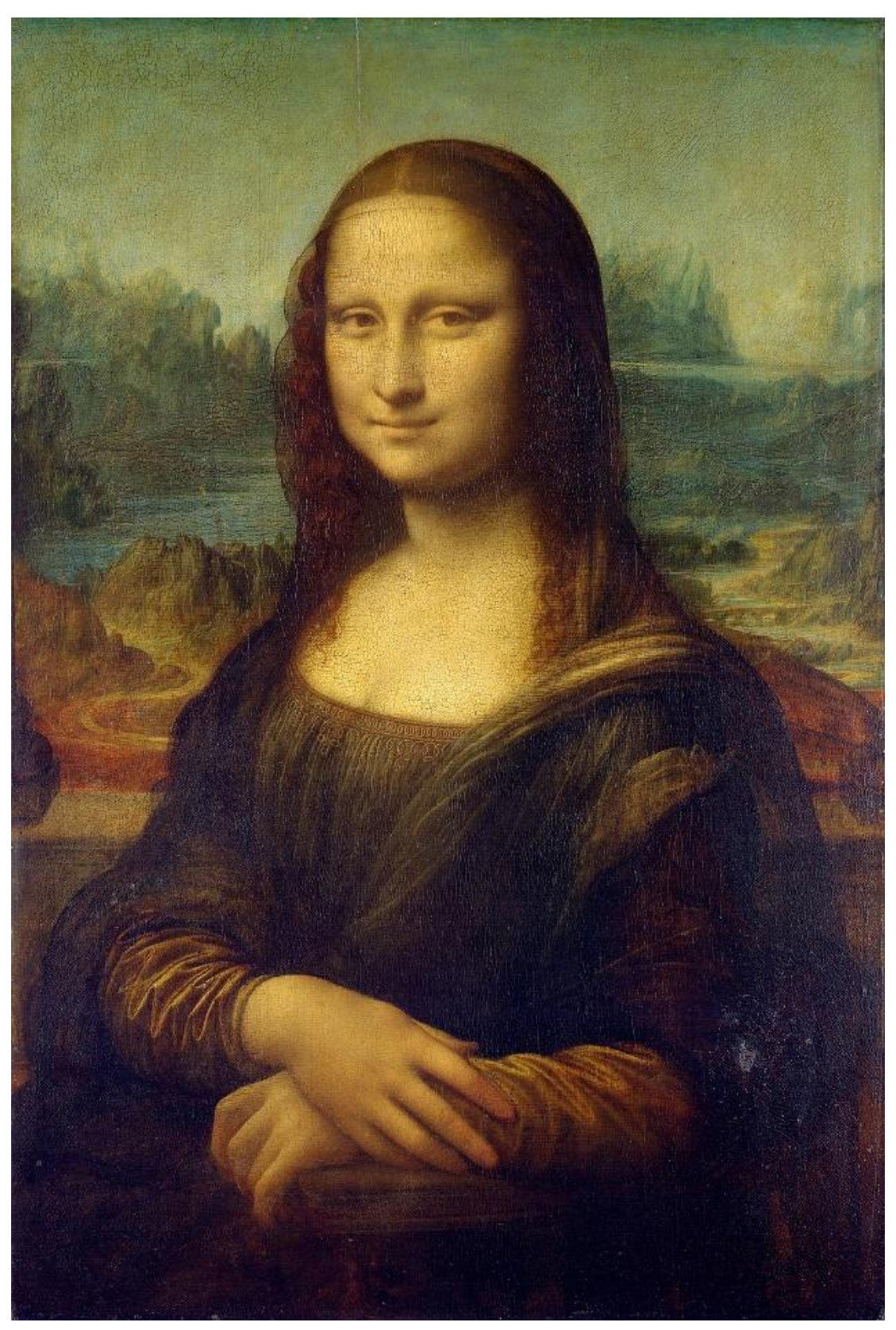

**Figure 2.1.** Portrait of Mona Lisa del Giocondo by Leonardo da Vinci.

Consider the famous Mona Lisa, painted by Leonardo da Vinci (Figure 2.1). What is the meaning behind her enigmatic expression? Is she pleased, or is it a sardonic grin? Does it signify a joke, flirtation, or some other hidden message? Often, viewers gaze upon this painting with bewilderment, and ultimately, each person walks away with their interpretation of this iconic and ambiguous artwork. In other words, every visitor to the Louvre Museum will perceive Mona Lisa's smile in their own unique way and react and comment accordingly.

This concept is illustrated by Plato's allegory of The Cave, which describes a group of people who have spent their entire lives chained to a wall in a cave, facing a blank surface. They could only see shadows of objects passing behind them and began naming these shadows. However, while the shadow of a horse was the prisoners' reality, it is evident that this is not an accurate representation of what a horse would look like outside the cave. This story implies that the reality outside the cave is more significant than the one inside. However, while discussing change, we will not make such a judgment and will consider all representations of reality to have the same dignity. The crucial point here is that convincing a prisoner that a horse is vastly different from its shadow would be challenging. We could go inside the cave, get to know the prisoners, and gradually persuade them to discard their chains and explore the outside world. This means we must embrace our neighbor's reality and convey change from within it.

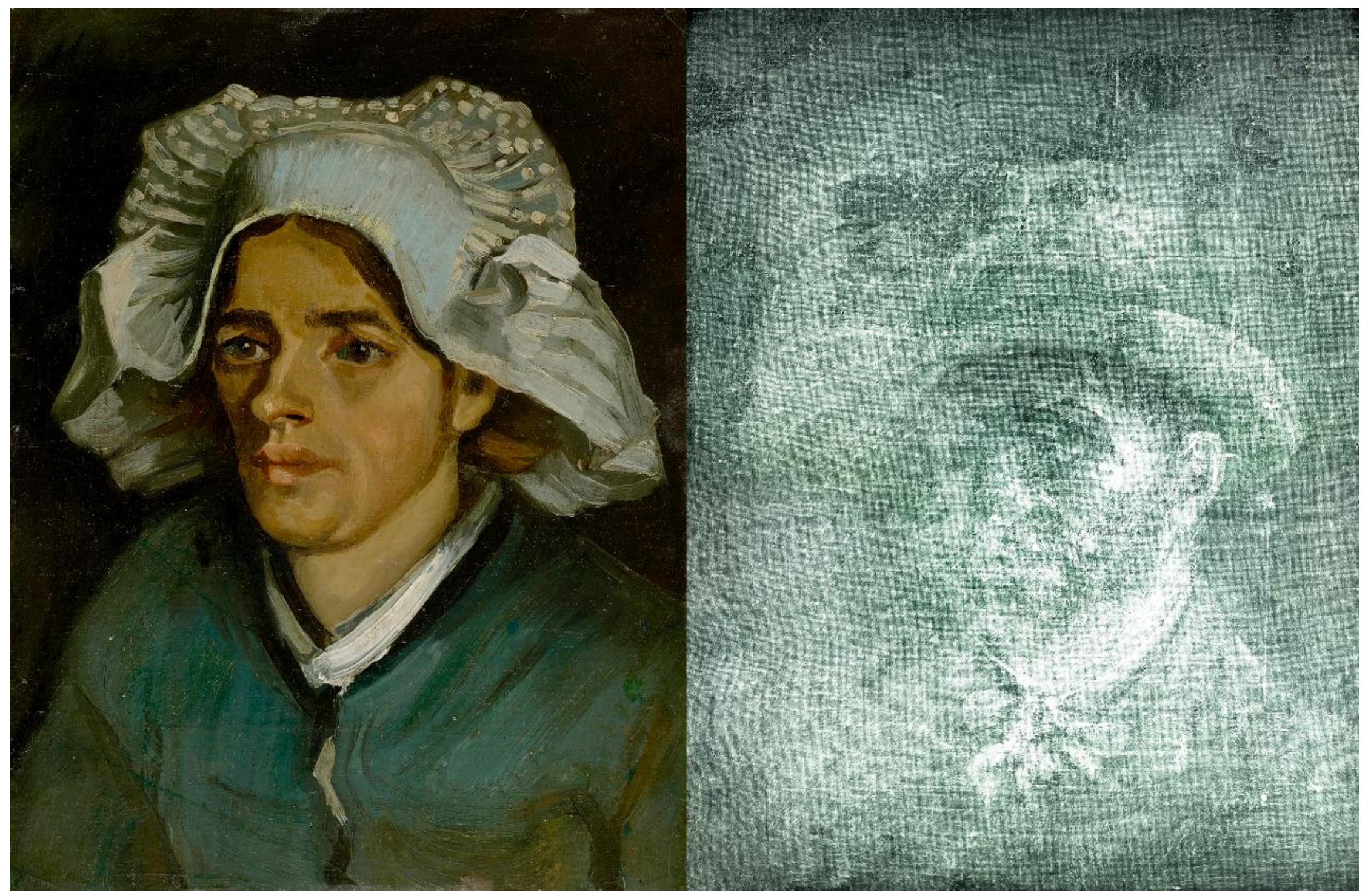

**Figure 2.2**. Vincent van Gogh's 1885 painting Head of a Peasant. Left: diffused light. Right: X-radiography.

As the last example, consider the x-radiography of Vincent van Gogh's 1885 painting Head of a Peasant (Figure 2.2). Imagine two observers examining the painting, one under diffused visible light and the other using X-rays. Which of the two would be more easily convinced that this is a self-portrait of the artist?

Putting ourselves in the shoes of others can be an excellent exercise to expand our perspectives. Sometimes, we may be stuck in our view, and our perception of reality can become a limitation. By asking questions like, "*If I had asked your sister (or partner or colleagues) their perspective on your problem, what would they have told me?*" we can gain new insights and find new ways to approach a situation.

A helpful exercise to begin with is to complete the Table 2.1 form. After stating a problem, we can describe how we perceive it and what our initial reaction would be. It is important to not only consider our perspective but also to think about at least five other people, including those close to us and those who may have different viewpoints. The exercise can be approached in two ways. The first approach, as shown in the table, involves considering how others might experience the same problem – exploring their perceptions, reactions, and underlying responses to similar situations. The second approach is to imagine or ask how others, observing us from an outside perspective, might interpret our problem or behaviors. By putting ourselves in their shoes, we can imagine their perceptions and subsequent reactions. For example, let us answer the question, what would my mother think of the situation? What factors would she emphasize? What would her immediate reaction be? And what about the long-term impact? Finally, let us evaluate these reactions, understanding whether they would be more rational or emotional. This initial step is critical to understanding how we perceive a problem or situation around a goal. The exercise also reminds us that our immediate reaction is never the only possible one.

At this stage, we do not need to worry about defining the problem in the best possible way. Instead, let us focus on writing the simplest and most immediate statement that comes to mind. In the following sections, we will explore how accurately identifying a goal or problem is the foundation for any change management intervention.

| PROBLEM STATEMENT | PERSON | PERCEPTION | REACTION | DOMINANCE (MORE EMOTIONAL OR RATIONAL) |
|---|---|---|---|---|
| I'M ALWAYS LATE FOR WORK | ME | I'M IN A VERY DEEP SLEEP AND I DON'T HEAR THE ALARM CLOCK | I WILL TRY TO BUY A LOUDER ALARM CLOCK OR ASK MY HUSBAND IF HE CAN WAKE ME UP | RATIONAL |
| | MY SISTER | I REALLY DON'T LIKE MY BOSS. I WISH I COULD CHANGE MY JOB | I HAVE A HARD TIME GETTING UP, AND I'M SLOW IN THE MORNING BECAUSE I'M DOWN IN THE DUMPS. | EMOTIONAL |
| | MY FATHER | I AM RETIRED. I GET BORED AT HOME. I WISH I COULD RETURN TO THE TASKS I PREVIOUSLY EXPERIENCED ONLY AS A CAUSE OF STRESS | I SPEND TOO MUCH TIME AT HOME, GET DEPRESSED AND DON'T HAVE THE STRENGTH TO DO ANYTHING | EMOTIONAL |
| | MY COLLEAGUE | | | |
| | A PERSON I ADMIRE | | | |
| | A PERSON I HATE | | | |

**Table 2.1**. Considering different perspectives.

## 2.1. The pace of change

When it comes to implementing change, the pace at which it occurs is a crucial factor to consider. *Gradual* change is often preferable when emotional resistance is minimal and an individual is willing to collaborate in implementing change (Nardone, 2016). For example, breaking down a complex task into smaller work packages can make it more manageable. However, not all changes can be gradual. Imagine what would happen if some citizens of London gradually switched to right-hand drive. In such cases, self-organization rarely works, and a central authority is needed to coordinate a simultaneous systemic change.

Conversely, *exponential* change is a shift in human behavior that starts slowly and then accelerates rapidly. It is like a snowball rolling down a mountain slope and getting bigger and faster. This type of change can be induced by small actions (even nudges[1]) that trigger a series of events, making it useful when there is a high initial resistance due to emotional barriers.

Consider the inspiring story of Paolo, a music lover who always felt awkward and uncomfortable while dancing. He was so ashamed of his lack of skills that he would make excuses whenever his friends invited him to a disco. However, everything changed when he started dancing for his two-year-old niece, Clara. With her, he felt free to let go and have fun without fear of judgment. This experience sparked a newfound love for dancing, and he began practicing in the privacy of his room. Today, Paolo is a renowned professional dancer, whose journey to success started with a small step.

On the other hand, *catastrophic* change is something we all want to avoid. It can hit us like a tsunami, leaving us feeling overwhelmed and helpless. We often receive warning signs that this kind of change is coming, but we choose to ignore them. This is where the Tarot card of the Tower comes into play (Figure 2.3). While it is often interpreted as a symbol of danger, crisis, or failure, it should be better understood as an opportunity for renewal and personal growth. A crisis can serve as a catalyst for change. The flame that descends from the sky in the card reveals the tower's roof, removing what was stagnant and creating space for transformation. By dismantling superstructures, we allow ourselves to breathe in fresh air. Contrary to the depiction in Rider Waite's tarot deck, the jesters in the card (Lequart Tarot deck) do not tragically fall from the tower; instead, they dance, play, and revel in the dynamism of change. This card imparts the wisdom that change is inevitable, and we can choose whether to embrace it and dance along or get overwhelmed by it.

[1] See the book of Thaler and Sunstein (2008) and later in this book for a better discussion of the power of nudges.

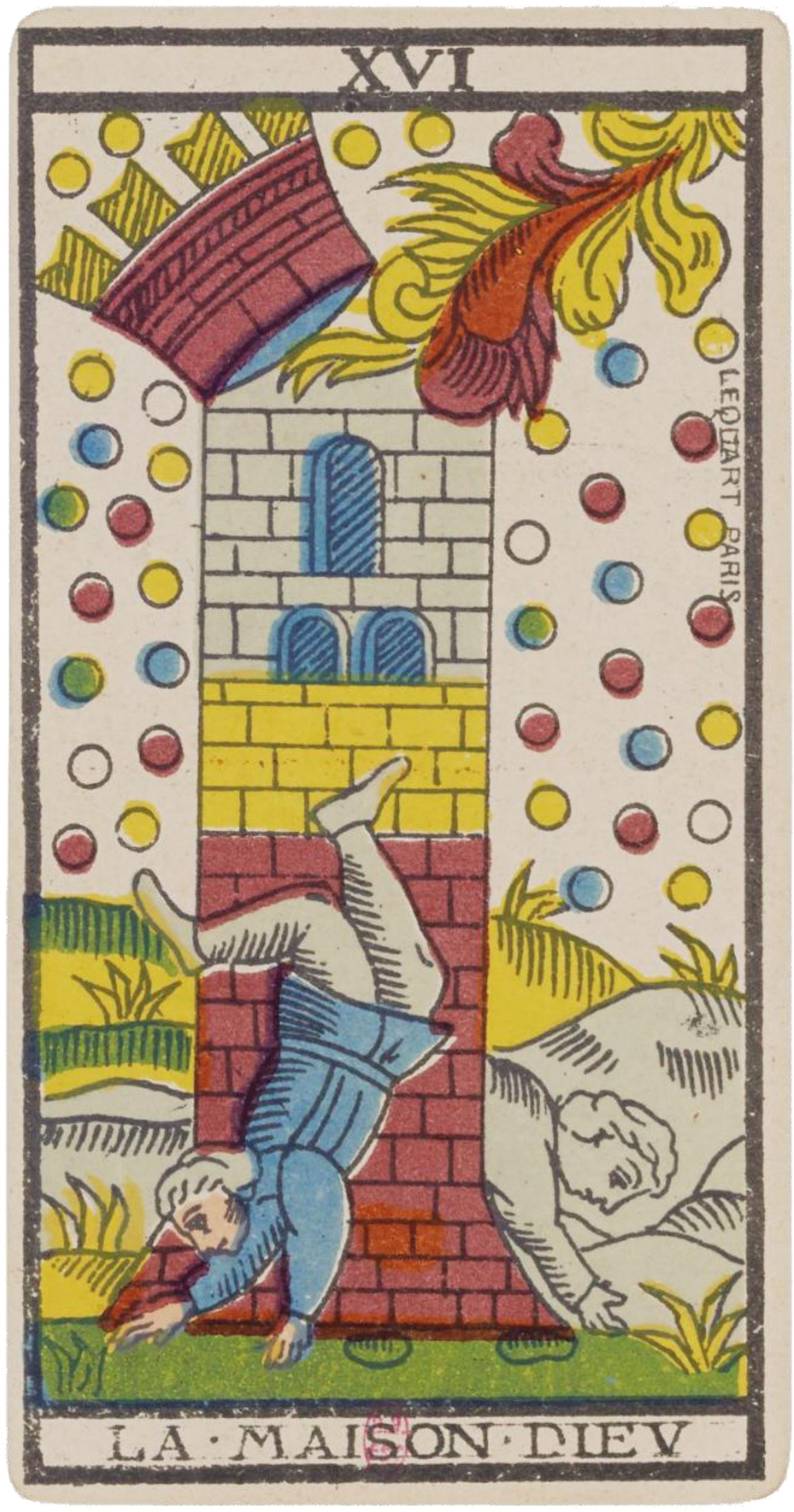

**Figure 2.3**. Tower – Lequart Tarot deck

In the following sections of this and the next chapter, we discuss the essential steps of a guided organizational and personal change. We examine problems that can be resolved through conventional logic, as well as those that stem from emotional barriers and require *corrective emotional experiences*, often involving unconventional approaches. Our discussion is also influenced by the works of Paul Watzlavik, Giorgio Nardone, and their colleagues (e.g., Nardone, 2009; Nardone & Balbi, 2017; Watzlawick, Bavelas, et al., 2011; Watzlawick, Weakland, et al., 2011).

### 2.2. Goal setting

The first step toward achieving change is identifying the problem that needs to be addressed or the goal that we want to achieve. This is crucial as it provides a clear sense of direction and helps individuals and teams understand where they want to go.

Committing to a goal is much easier than making a broad commitment, and change often requires a significant effort from everyone involved. This effort is not just rational but also requires challenging the status quo and accepting a certain degree of risk. As we would look for a "reason why" the consumer should buy our products, here we look for a motivation for people or employees to change.

Very often, personal growth and team performance are hindered by not seeing a clear destination ahead – so that if one does not know to which port one is sailing, no wind is favorable[2]. Therefore, it is important to help individuals visualize the scenario they want to reach. A typical question we would ask is:

"*If tomorrow morning, **by some miracle**, you woke up and your problem was solved, what would happen? What would you **do differently** from when you wake up to when you go to sleep? What **new emotions** would you experience? What would make you think the problem has been solved?*".

Better if we can ask contextual questions about the specific case. If a person is suffering from mobbing in the workplace, we would ask questions such as:

"*What would change **in your relationship** with colleagues if by miracle the problem was solved? What are the **new emotions** you would experience while going to work? What would be the **change indicators** while you are in the office?"*

The response from a person involved in a case we worked on was:

*"The office environment has significantly improved, with a noticeable decrease in the negativity that once characterized interactions. Instead of dealing with tension, I now see friendly greetings and positive interactions.*

*The emotions I feel on my way to work have changed from anxiety to optimism. There's now a genuine sense of excitement, knowing that the day will involve constructive collaboration. The stress that used to weigh me down has been replaced by a feeling of empowerment and belonging.*

*These positive changes are evident in the team dynamics: there's more open communication, laughter, and idea-sharing. Team members actively support one another, creating a*

---

[2] Seneca. Epistolae, LXXI., 3.

*collaborative and innovative atmosphere. Meetings have shifted from tense exchanges to productive, solution-focused discussions.*

*The scars of the past are healed."*

Emotional blocks can prevent individuals from picturing their final goal or finding a solution to their problems. Fear of change or its consequences can be a significant obstacle. For instance, a manager promoted to their dream position may have to move abroad and be separated from their family. In such cases, individuals may leave the job and return to their previous position, as they cannot bear the side effects of change.

For this reason, visualizing a scenario that seems impossible but could miraculously come true is crucial in freeing the mind from emotional blocks and encouraging unconventional thinking. It enables individuals to explore ideas that may seem unattainable in their reality and gain insight into the underlying emotions associated with the problem. To enhance this process, it is even more effective to accompany the scenario with vivid visualizations of the narrated scenes. The person should be immersed in a vivid experience where they can truly live and emotionally connect with the imagined scenes.

Helping people visualize a positive scenario can move their vision beyond obstacles and create a *self-fulfilling* prophecy. When individuals foresee a positive outcome, a goal that has been reached, they are motivated to act. This kind of visualization can introduce new beliefs and expectations and lead to a shift in behavior, even unconsciously. For example, if we believe that people will admire us at a conference, we will dress nicely, smile, and prepare the best speech. This positive mindset can lead to admiration from others. Conversely, if we believe that everyone hates us, we may give the worst of ourselves in that situation. Self-fulfilling prophecies are based on our beliefs and perceptions of reality. If we perceive a possible scenario to be real, it will produce real consequences (Merton, 1948).

Effective change also requires clear benchmarks. We must identify *what* we aim to improve and the *timeline* for achieving it. *Measuring* progress is crucial, as it allows us to understand the distance between our current state and the desired outcome. While this distance may not always be objectively measurable, individuals must become aware of it to make meaningful progress. One useful tool for self-assessment is the zero-to-ten scale. By asking individuals to rate their current position on this scale and identify the smallest next step that would move them forward, we can create a roadmap for progress. We could ask them:

*"If we wanted to **measure** where you are now on a scale of zero to ten, what number would you tell me? What is the **smallest next step** that would move you forward by one point?"*

In summary, when initiating an individual or group growth process, the first step is to set a relevant goal, make it specific and measurable, but most importantly, to visualize future scenarios and make them real and possible in the perception of the people who want to change. The failures of change often lie in the fact that goals are imposed. Instead, people should be helped to find and visualize their *own* goals.

The form presented in Table 2.2 serves as a tool for summarizing the findings obtained through the miracle scenario technique. In the initial section, we will outline the emotions and key elements of the envisioned scenario, emphasizing the tangible differences from the current situation. Subsequently, we will proceed to determine the timing of the change and its concrete indicators, which are the factors that truly signify a new situation. We will assess our current position on a scale of one to ten in relation to achieving our objective. This assessment will serve as a preliminary step in defining the subsequent incremental actions required to bring about tangible change. If the scenario can connect well to a self-fulfilling prophecy, these small steps may involve actions that simulate the problem's resolution. One should act *as if* the problem were solved. For example, if we perceive a colleague as antagonistic, we could practice daily acts of kindness toward them, treating them as an ally rather than an adversary. We better discuss this concept later on.

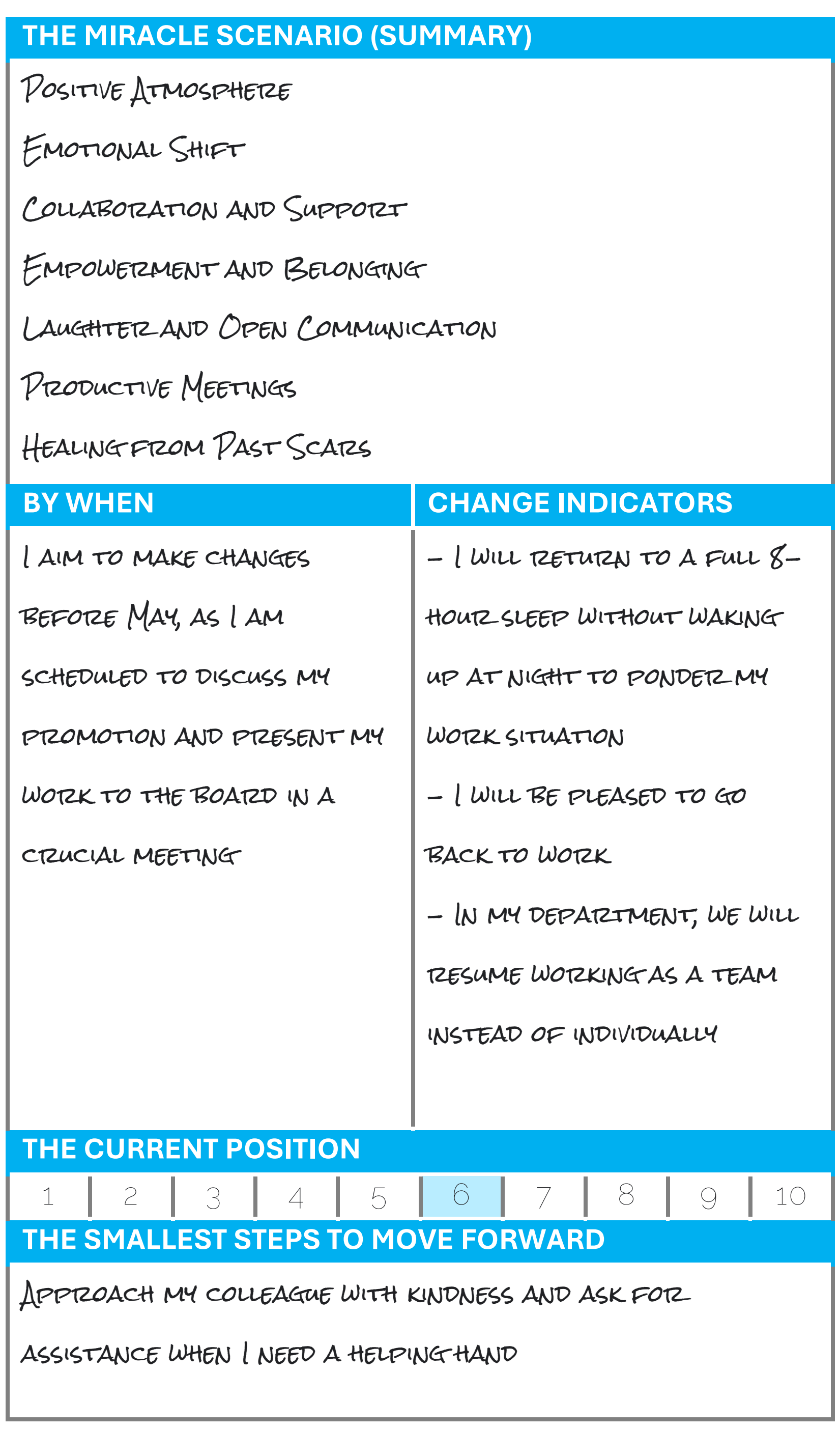

| THE MIRACLE SCENARIO (SUMMARY) | |
| --- | --- |
| Positive Atmosphere | |
| Emotional Shift | |
| Collaboration and Support | |
| Empowerment and Belonging | |
| Laughter and Open Communication | |
| Productive Meetings | |
| Healing from Past Scars | |
| **BY WHEN** | **CHANGE INDICATORS** |
| I aim to make changes before May, as I am scheduled to discuss my promotion and present my work to the board in a crucial meeting | – I will return to a full 8-hour sleep without waking up at night to ponder my work situation<br>– I will be pleased to go back to work<br>– In my department, we will resume working as a team instead of individually |

| THE CURRENT POSITION | | | | | | | | | |
| --- | --- | --- | --- | --- | --- | --- | --- | --- | --- |
| 1 | 2 | 3 | 4 | 5 | **6** | 7 | 8 | 9 | 10 |

| THE SMALLEST STEPS TO MOVE FORWARD |
| --- |
| Approach my colleague with kindness and ask for assistance when I need a helping hand |

**Table 2.2**. The miracle scenario.

## 2.3. Understanding what is jammed

Once a goal has been set, the first step towards achieving it is understanding why it has not yet been reached. This involves examining all the *attempted solutions* and actions taken by individuals or teams thus far. It is crucial to investigate *who* did *what* and with what *results*.

It is important to note that the purpose of this search is not to find someone to blame for their actions. Instead, we would ban the word *"why"* from our questions, as it often leads us to believe that there is a *linear cause* for every problem and that someone must be at fault if something does not work. In our questions, we should replace "why" with *"how does it work,"* even if it seems unnatural. For example, instead of asking:

"**why** did we not reach the number of sales we had planned?"

we would ask,

"**how does it work** that we did not reach the number of sales we had planned?".

The focus must shift from searching for a single cause to analyzing the mechanisms fueling a problem. If someone has made a mistake, it is essential to understand what led them to make that mistake and how to prevent it from happening again. In ordinary *cause-and-effect logic*, we tend to think that one event causes another, often following a *linear path*. However, the formation of problems is often much more complex than it appears. A problem, or even a positive outcome, originates from the *complex interaction* of various elements in a system. We deal with complex systems where the properties of the *whole*, i.e., the final result originating from the actions of multiple actors, cannot be understood just by adding the system components.

Complex systems are dynamic and ever-changing, exhibiting long-term behaviors that can be unpredictable. Take, for example, schools of fish. These creatures swim in harmony, avoiding collisions with one another without any central authority or fish to coordinate their movements. Instead, each fish instinctively follows small rules that keep it at a safe distance from its peers. The collective movement of the school *emerges* from the unplanned behavior of individuals, making the system robust and less susceptible to predator attacks.

Similarly, two individuals may possess exceptional intelligence when working independently. However, if they harbor animosity toward one another, their collaboration within an innovation team could detrimentally impact the overall performance, resulting in subpar outcomes. Conversely, two employees who individually yield unsatisfactory results may synergistically complement each other, thereby forming the most innovative team within their organization.

It is important to understand that the *whole is something besides the parts*[3]. When trying to solve a problem, looking for a culprit is not only useless but often harmful. This way of thinking moves us away from solving the problem itself. Instead, we must focus on understanding the mechanisms that create the problem and interrupting any actions that might worsen it, even when they appear as a solution. To do this, we must carefully analyze attempted solutions, evaluate their consequences, and distinguish between harmful and harmless actions.

[3] Aristotle, Metaphysics, Book VIII, 1045a.

People often make mistakes in their actions, either by *acting when they should not* (such as blurting something out in anger during a meeting or speaking before having listened to another's proposal in a negotiation), *not acting* when they should (such as not addressing hurtful behavior in a relationship), or *acting in the wrong way* (such as making a promise or a threat that is not followed through).

*Repetitive* attempts to solve problems can backfire by turning into automatic responses that may no longer work or even become harmful habits. We may be tempted to rely on solutions that worked before or in other situations, but we must be aware that the context and conditions may have changed. For example, driving to work during a flood is not viable.

To become more mindful of our problem-solving behavior, we should explore *what* happens to *whom*, *when,* and *where* without neglecting any details. We should also reflect on whether we have tried anything before, how we react when the problem arises, whether we face or avoid it, and whether we seek help or go solo.

It is crucial to comprehend whether our actions are deliberate and voluntary or if they arise spontaneously, only becoming apparent after they have already begun. There is a significant difference between punishing a child to educate them and slapping them out of anger. Similarly, if we find ourselves shouting at colleagues to defend our opinions in a meeting, we risk being on the wrong side. However, our impulses are often challenging to control or comprehend. Sometimes, we act because we feel compelled to do so due to *external* or *internal pressures*.

Consider the case of a student from a family of engineers who wants to study psychology. They may feel pressured to choose an engineering degree to avoid disappointing their parents, leading to a sense of burden that can hinder their exam preparation.

It is essential to consider the consequences of our attempted solutions. How do they impact our future behavior and emotions? How do they affect individual and team performance? We must recognize actions that exacerbate problems instead of solving them. These actions may *not work*, *may only partially solve the problem*, or *may have worked in the past but are no longer effective*.

Dysfunctional solutions may become habits and hinder our ability to find better alternatives. Sometimes, we may stumble upon a great solution but lack the necessary skills or emotional fortitude to *implement it*. Other times, a solution that once worked well may become dysfunctional due to our *inability to maintain* it or its unforeseen *side effects* (Milanese & Mordazzi, 2014). This could be the case of a leader who feels overworked and starts delegating some of his tasks, but being a control freak, he cannot bear to lose control. Similarly, one might wisely quit smoking but fail to cope with the increase in body weight (nicotine increases resting metabolic rate), ending up returning to cigarettes.

Repetitive habits can *cloud our perceptions* and interpretations of external stimuli and lead to self-deception or intellectual blindness (Bach, 1981). When our perceptions change, our reactions will also be distorted. We may become unable to *react* or *not react appropriately*.

Once we recognize a dysfunctional solution, the most powerful thing we can do is *interrupt it*. In most cases, this may already guide us toward a new solution and generate new perspectives to solve our problems. Indeed, if we truly want to know how something works, we should try to change it[4].

Identifying and *disrupting dysfunctional solutions* is a crucial step after understanding our goal (Watzlawick, Weakland, et al., 2011). All the theories of change and practical advice we will see in this book can connect to this concept. These strategies involve leveraging emotion (or reason) to disrupt dysfunctional solutions or *replicate* successful ones. However, it is important to note that understanding harmful behaviors is only the first step and does not always lead to stopping them. Some problems can only be solved by *changing perceptions* and consequent reactions, through *a new way of feeling*.

A key characteristic of dysfunctional solutions is their repetitive and persistent nature. Trying an approach that does not work is not inherently wrong; the issue arises when we repeatedly turn to the same ineffective behaviors, which ultimately reinforce the problem itself. Accordingly, it is essential to recognize that there is nothing inherently flawed in our attempts to act or refrain from action while seeking a solution. It is natural to realize that a particular approach is ineffective and subsequently change course. In such cases, the behavior cannot be deemed dysfunctional. Almost all innovation processes rely on *trial-and-error* methodologies. However, problems arise when we stubbornly persist in pursuing a direction that has already proven unsuccessful.

Table 2.3 is a valuable tool for change managers and individuals seeking to analyze their attempted solutions and identify dysfunctional ones. The diagram in the table guides a detailed analysis of their characteristics and possible consequences. It helps us understand which solutions are ineffective and need to be interrupted.

On the other hand, when we come across exceptions (see Table 2.4) – solutions that have proven successful in the past – our first attempt should be to replicate them, if still possible and applicable to the current context.

[4] Quote by Kurt Lewin.

| GOAL/PROBLEM (BRIEF RECAP) | | | | |
| --- | --- | --- | --- | --- |
| Being able to present his work in English at company meetings without fear | | | | |
| **ATTEMPTED SOLUTION (REPETITIVE, CAN BE MORE THAN ONE)** | | | | |
| **WHAT** | **WHOM** | **WHERE** | **WHEN** | **HOW** |
| He is afraid of speaking in English because it is not his mother tongue. He calls him in sick or asks his colleague to present in his place. | It involves him and his colleagues. Instead, he has no problems speaking English with his friends from UK | At the office | During company meetings. This does not happen if he is alone with his boss. | He feels completely stuck, can't remember what to say, his voice is shaking. He feels the need to avoid the presentation. |

| | | | |
| --- | --- | --- | --- |
| **CLOUDED PERCEPTION** | He thinks avoiding the presentation or having someone else present will save him from making a bad impression. He fails to realize that every time he asks for help or avoids a situation that frightens him, it reinforces his sense of incapacity. | | |
| **INABILITY** | **CANNOT FIND A (DIFFERENT) SOLUTION** | CANNOT APPLY THE SOLUTION FOUND | CANNOT STAY THE COURSE OR HANDLE SIDE EFFECTS |
| **TRIGGER** | SUFFER THE ACTION. CAN NOT HELP BUT DO IT OR NOT DO IT | **STARTS BY DELIBERATE CHOICE** | IT TRIGGERS SPONTANEOUSLY. EITHER NOTICE IT OR CONTROL IT WHEN IT HAS ALREADY STARTED |
| **EFFECT** | ACT WHEN THEY SHOULD NOT | DO NOT ACT WHEN THEY SHOULD | **DO THE WRONG THING** |
| **WHY DYSFUNCTIONAL** | **DOES NOT WORK OR THINGS GO WORSE** | WORKS ONLY PARTIALLY | ONLY WORKED IN THE PAST OR IN A DIFFERENT SITUATION |

**Table 2.3**. Analysis of attempted solutions (use this table more than once, in case of several attempted solutions).

The example presented in Table 2.3 illustrates the case of a collaborator who struggles with delivering presentations in English during business meetings. The table provides an analysis of two attempted solutions – seeking assistance from a colleague and avoiding presentations by feigning illness – which could be expanded to include additional attempted solutions if necessary. Upon initial examination, it becomes evident that the problem only arises within the office environment and not when conversing in English with friends. The employee experiences uncontrollable physical reactions due to fear, leading him to resort to avoidance as a solution, either by seeking help or making excuses. Although avoidance initially provides a sense of security, it subconsciously reinforces his fear and feelings of inadequacy.

The second part of the table examines the triggers, effects, and underlying inabilities associated with the attempted solution. In this case, asking for help or avoiding presentations are conscious choices the individual makes. They are not spontaneous reactions that become conscious only after they have commenced, nor are they involuntary actions resulting from emotional dynamics. Fear hinders a person from believing in their ability to overcome the challenge and explore alternative solutions. Consequently, the employee deliberately makes a detrimental choice that further reinforces his sense of incapacity. The situation would have been different if he had chosen to confront his fear by presenting at company meetings, even if he froze when faced with the first question from colleagues.

## 2.4. Guiding a predominantly rational change

In some cases, change can be primarily rational and achieved by following an almost *ordinary logic*. These are the cases where individuals have both the skills and the emotional resources to solve a problem. In such cases, our initial step in driving change should be to look for *exceptions* and replicate them if still possible and applicable to the current context.

*What has worked in the past to solve the same or similar problems?*

If someone is stuck in finding a solution, we can explore the behaviors that have helped them in the past. We must also bear in mind that our actions occur on three levels: *communication* (what we say, how we say it, and our nonverbal expressions), *thoughts* (our internal dialogue and self-talk), and *actions* (including the decision to take no action). Remember that every action or inaction is a choice, which can be more rational or emotional, conscious or unconscious. When we find exceptions, we must evaluate whether they can be repeated. Something that worked in the past is not guaranteed to work now. The actors involved may have changed, or the person may be unable to replicate the exception. The context may have changed.

Consider the scenario of a student struggling to pass an exam while living in a chaotic and crowded household with three young siblings vying for their attention. Finding a quiet space to concentrate in such an environment can be daunting. However, there is one exam that the student passes with flying colors. How did they manage to do it? Was it a one-time fluke or a repeatable exception? If the student could ace the exam by copying answers from their best friend, this exception is hard to replicate. However, if the student could spend a few days before the exam alone at a family lake house, with peace and quiet to concentrate, this could be a repeatable exception. Identifying and replicating exceptions is a crucial step in guiding change and solving problems.

Table 2.4 serves as a tool to extend our previous example (see Table 2.3) by identifying potential exceptions. This table presents two illustrative solutions that have proven effective in the past but are not applicable or useful in the current context.

The first example makes us realize that the most feared colleague is the native English speaker, primarily due to their exceptional command of the language. One could consider fostering a positive relationship with this colleague to address this issue, transforming them from a judge into an ally. Another option might be to rely on meetings where this colleague is not present. However, it is important to note that the employee has no control over their presence in meetings, rendering this option unrepeatable in immediate circumstances.

The second solution, resorting to anti-anxiety medication, is a path that, in most cases, should not be revisited. Even if one were to find a doctor willing to renew the prescription, long-term reliance on such medication would only reinforce a sense of dependence without truly resolving the underlying problem. Imagine forgetting the medication on a crucial day or finding oneself in a location where it is unavailable. Relying on medication once again becomes a plea for external assistance, conveying the message that one is incapable of managing the situation without it.

| EXCEPTIONS (SOLUTIONS THAT WORKED IN THE PAST) | CAN THE SOLUTION BE REPEATED NOW? | WOULD THE SOLUTION WORK? |
|---|---|---|
| 1. HE GAVE AN EXCELLENT PRESENTATION ON THE DAY HIS COLLEAGUE FROM THE UK WAS NOT PRESENT AT THE MEETING. | NO | YES |
| 2. HE DID A GOOD PRESENTATION THE DAY HE TOOK ANXIETY MEDICATION. | YES | NO |
| 3. ... | | |

**Table 2.4**. Analysis of exceptions.

In many cases, people tend to settle for the first solution that comes to mind, even if it may not be the best option. This is especially true for complex and interconnected problems that lack clear definitions. Therefore, investing time in the goal-setting phase is crucial to clarify the final objective, which may differ from the initial idea. Often, people struggle to identify their problem or goal or overlook critical constraints. When individuals or teams come up with a quick solution for a complex problem that leads to suboptimal results, it is essential to guide them toward exploring *alternative solutions*.

Take the problem of global warming, for example. While planting trees is a popular solution, it is not the only one. Trees effectively remove carbon dioxide from the atmosphere and provide shade, but other ways exist to tackle this problem. To expand our solution space, we can use well-known techniques such as brainstorming, mind maps, or the six thinking hats[5] (Buzan, 2010; De Bono, 1985). These methods can help us generate additional points of view and solutions. For instance, reducing meat consumption and switching to plant-based alternatives can significantly reduce our carbon footprint. The production of meat, especially beef and lamb, requires many resources, including land, water, and energy. Moreover, meat production generates significant greenhouse gases, primarily methane. Another effective solution is transitioning to clean energy. Fossil fuels such as coal and oil are the primary sources of greenhouse gas emissions. By reducing our reliance on these fuels and switching to renewable energy sources, we can significantly mitigate the effects of global warming. In conclusion, while planting trees is a valuable solution, it is not the only one.

Lastly, when faced with a complex problem, it can be advantageous to break it down into smaller, more manageable tasks. This approach can help to alleviate the anxiety that often accompanies tackling a large and daunting challenge. Therefore, those responsible for driving change must guide individuals toward achieving their goals by breaking them down into smaller, more achievable tasks. For instance, if one were to embark on building a magnificent castle, one would first research historical sites and gather visual inspiration, define the style, develop a detailed floor plan, prepare the site, build the walls, and so on. While the overall task may seem overwhelming, breaking it into smaller, more manageable sub-tasks can make it feasible. The more stages, the easier it is to plan the steps needed to reach the goal. Moreover, it is essential to *plan the route from the summit going down*, much like a climber would when planning their ascent of a challenging mountain. This means starting by considering the goal reached and then planning the step immediately before it. This approach is highly effective as it prevents individuals from getting lost on the wrong path, leading them away from the summit. Beginning to think of the summit as already reached can also create a self-fulfilling prophecy.

We present an example in Table 2.5, exploring the goal of creating a sophisticated humanoid robot that can comprehend and perform complex human tasks. We have broken down the initial problem into ten sub-tasks, but it is important to note that each can be further divided. The goal is to turn the original problem into a series of small steps that provide guidance and do not frighten team members. As you will observe, the tasks have been defined in reverse order, starting from the last and concluding with the first.

---

[5] This is a decision-making and problem-solving method developed by de Bono (1985). It involves six methaphorical "hats," each representing a different type of thinking: white (facts and information), red (emotions and intuition), black (caution and critical judgment), yellow (optimism and benefits), green (creativity and new ideas), and blue (organization and process control). With different team members focusing on one perspective at a time, it helps groups explore ideas thoroughly and make wise or innovative decisions.

| MAIN GOAL |
| --- |
| Creating a sophisticated humanoid robot that can comprehend and perform complex human tasks. |
| **TASKS** |
| 10. Conduct extensive testing to validate the reliability and functionality of the robot. |
| 9. Achieve full autonomy by refining and optimizing the entire hardware infrastructure to support the execution of complex tasks independently. |
| 8. Integrate all subsystems into a cohesive humanoid robot system |
| 7. Develop natural language processing and communication abilities for effective interaction with humans. |
| 6. Implement algorithms that enable the robot to analyze complex situations and solve problems autonomously. |
| 5. Integrate a system for the robot to understand and respond appropriately to human emotions. |
| 4. Develop machine learning capabilities to enable the robot to learn from experiences and adapt to new tasks. |
| 3. Build a robust mechanical and electronic system, including motors and sensors, for efficient navigation and locomotion. |
| 2. Design and integrate cameras, grippers, and actuators for object identification and manipulation. |
| 1. Develop sensors for recognizing and responding to human gestures and speech. |

**Table 2.5**. Problem breakdown.

## 2.5. Unconventional logics

By analyzing the attempted solutions to a problem, we can determine whether a rational or emotional approach is more effective in organizing the path to change. In certain situations, individuals may possess the technical skills necessary to solve a problem. However, they may struggle with the emotions surrounding the issue or the potential consequences of their actions. Even if they comprehend what needs to be done and how to do it, they may be emotionally blocked, preventing them from taking action or leading their actions in the wrong direction.

For example, a researcher may have the opportunity to join a renowned research team abroad for three months, which would undoubtedly be a significant career boost. However, they may still miss out on this opportunity because they fear leaving their comfort zone or their partner behind for that period. Without emotional and practical support, the researcher may choose to remain in their home country. When making difficult decisions, it is sometimes easier to postpone them until it is too late, resulting in *inaction* becoming a choice. Accordingly, understanding the emotional and rational aspects of problem-solving is critical in determining the most effective approach to change. By recognizing and addressing emotional barriers, individuals can overcome obstacles and make informed decisions that lead to positive outcomes.

Indeed, our repetitive behavioral patterns can stem from emotional barriers or distorted perceptions, leading to unfavorable actions. While we believe in the individual perception of reality and respect that, it is important to acknowledge that certain states of mind can result in more productive responses, ultimately fostering better communication and relationships. Those who struggle with emotional or perceptual obstacles often experience subpar performance in personal and professional settings. In such cases, simply relying on *cognitive* solutions may not suffice. While rational understanding and self-awareness are crucial steps toward improvement, they may not always be enough to change behavior.

Consider the case of our friend who had a deep-seated fear of flying. Despite undergoing three years of psychotherapy, he could not overcome his fear until he discovered the root cause of his anxiety. As it turned out, his fear was triggered by a traumatic incident that occurred when he was just four years old - a fight with his father at the airport. He had repressed this memory, likely as a *defense mechanism* (A. Freud, 1992), until it was uncovered during therapy. While understanding the reason behind his fear was a crucial first step, it was not enough to solve the problem. Our friend needed to undergo a series of *emotional corrective experiences* that went beyond mere cognitive understanding. There was no point in telling him that “flying is secure, accidents are extremely rare and unlikely, and most people waiting for a flight are not scared”. Instead, he was guided to do two things. First, he was encouraged to visit the airport and observe people's behavior. He was told that “everyone experiences fear when flying” and that “airplanes are indeed a risky mode of transportation”.

By studying people's behavior, he looked for fear signals and could not find them. Our friend realized that most people were relaxed and enjoyed shopping at the duty-free store before their flight. A doubt crept into his mind, leading him to change his perception of reality, with no *truth* or explanation being imposed by somebody other than himself. This experience reduced his anxiety and gave him the courage to fly. Second, once onboard, he was asked to assume a painful position with his hands and keep it until landing. The pain caused by this exercise distracted him from the most frightening moments of departure and landing. This *shift of attention* proved to be a powerful solution to his fear. In addition to these exercises, our friend was advised to visit the airport daily, which had a desensitizing effect. Over time, his fear of flying dissipated, and he could travel without anxiety. While understanding the root cause of anxiety was important, it was not enough to solve the problem. This case highlights the importance of emotional corrective experiences. A combination of cognitive and emotional interventions is often necessary to produce change.

Clearly, a cause-and-effect logic is not enough to untangle emotional blocks. One should rely on unconventional logic, for example based on *paradoxes* and *contradictions* (Nardone & Balbi, 2017). A paradox is a situation in which two seemingly contradictory statements or conditions exist simultaneously, resulting in something that cannot be solved straightforwardly. According to Watzlawick, Weakland et al. (2011), it often occurs when people try to solve a problem using the same logic or approach that created the problem in the first place. The paradoxical nature of a problem becomes evident when our efforts to solve it actually worsen the situation, resulting in a vicious cycle of repeated unsuccessful attempts. This can be frustrating and confusing because there seems to be no way out. Paradoxes can also arise when different levels of communication or understanding clash with each other. For example, paradoxical communication can occur when a message is conveyed at one level of communication, but the recipient interprets it at another level. This can create a *double bind* (Bateson, 1972), in which a person receives, at the same time, two contradictory messages, one explicit and the other implicit. The person cannot escape the contradictory messages, and each of their responses risks being inadequate or wrong.

Consider the scenario of a manager who feels threatened by a highly capable but lower-ranking colleague. This manager may resort to constantly belittling their colleague's work. For example, when the colleague presents an essay they wrote, the manager may harshly criticize it but keep it on their desk instead of trashing it as usual. The latter is a gesture that indicates an appreciation for the work done. This behavior is an example of a double bind, where positive and negative signals are mixed, making it difficult to distinguish between them. Another classic example of a double bind is a mother who calls her son and expresses her desire to hug him, but when he runs up to her, she remains physically rigid and distant. Such situations can lead to anxiety and confusion, as it becomes challenging to interpret mixed signals.

Contradictions and paradoxes are two concepts that are often confused with each other. While they share some similarities, they differ in a crucial way. Contradictions involve conflicting true and false signals with a time gap between them. Paradoxes, on the other hand, involve contradictory statements that convey both true and false signals simultaneously. Contradictions are common in human behavior, yet many people are unaware of them. For example, we all know someone who wants to lose weight but refuses to exercise or a student who wants to get top grades but does not put in the effort to study.

So far, we have presented paradoxes and contradictions as something apparently negative. In the remaining part of this chapter, we explore how they can be used to break down emotional barriers and push us toward positive change. Lastly, we consider problems arising due to inflexibility, hard-to-change beliefs, and self-deception (Nardone & Balbi, 2017).

### 2.5.1. Contradictions

In today's society, there seems to be a growing fear of commitment to a single partner and the idea of being tied down to a relationship that limits one's options. People are drawn to the idea of fluidity and impermanence, as described by sociologist Zygmunt Bauman (2000). Additionally, there is a certain allure to the chase and the conquest, especially when it comes to dating and relationships.

Imagine someone who struggles with a lack of confidence and has difficulty presenting themselves in an attractive light on a first date. It may be that this person's lack of confidence causes them to seek constant confirmation or give too much of it after the first date, making the other person feel pressured. At times, introducing subtle contradictions and remaining partially veiled can add mystery and intrigue, helping to capture and hold attention more effectively. This could involve alternating between positive and negative messages, avoiding a clear vision. After a first date, one might express the pleasure of having met and spent time together and want to see each other again soon (thus providing positive feedback). Later, when the partner asks to see each other again, one might put this off by appearing very busy (thus sending a negative signal and making the other person feel not among their top priorities). We want to clarify that we are not recommending this behavior, but it can often result in temporarily binding the other person to us.

The logic of contradiction can be a valuable tool in guiding change interventions, for example, when dealing with fear or when we want to encourage people to learn new skills. Consider a scenario where a team is tasked with launching a new product. One team member is responsible for using innovative models to analyze consumer perceptions of their company's products and those of their competitors. This requires using software that the team member has never used before, which can lead to feelings of insecurity and a constant need for reassurance from their boss. The boss may observe the team member struggling to solve a problem for which they already have a solution. While stepping in and taking over the task may be tempting, doing so would have two negative effects. First, the boss would be bogged down in operational tasks and micro-problem resolution, taking away valuable time from other strategic activities. Second, the team member would miss out on learning and growth opportunities, becoming overly reliant on their boss for solutions. Furthermore, providing immediate solutions would stifle innovation in the execution of the task. By allowing the team member to work through the problem independently, they may develop a more innovative procedure and solution. Indeed, "if you tell people where to go, but not how to get there, you'll be amazed at the results"[6].

---

[6] Quote by George S. Patton.

A wise strategy for the boss would be to *observe* their employees *without intervening*, allowing them time to learn and grow. However, if an employee constantly asks for help, the boss should address their anxiety and restructure their perception of receiving help. It is important to guide the employee to understand that whenever they receive help from someone, they implicitly send them two messages: “I help you because I care” and “I help you because you cannot do it alone.” While the former is certainly positive, the latter is detrimental to the employee's ability to believe in themselves and develop their abilities.

In the next chapter, we also show how the logic of contradiction can be useful when dealing with individuals who oppose change. Now, we discuss some techniques that follow the logic of paradox.

### 2.5.2. Paradoxes

When a team or manager struggles to find a solution to a problem, it can be helpful to encourage them to think outside the box and consider alternative approaches. One effective method is encouraging them to change their perspective and consider how someone else, such as Steve Jobs or a competitor, might approach the issue. This technique, aligned with the idea behind the *six thinking hats* (De Bono, 1985), can help individuals view the problem from a different angle and generate new ideas. However, there are times when changing perspective alone is not enough to overcome a significant challenge. In these situations, paradoxes can be a powerful tool, and we would ask:

What could you ***do, think,*** *or* ***say*** to ***deliberately*** make the problem ***worse***?

Interestingly, individuals struggling with a problem often find it easier to identify actions that could worsen the problem. This realization can be incredibly powerful, as it helps them become aware of the factors contributing to the problem; some actions they put in place, probably as a dysfunctional solution (Nardone, 2009). Becoming aware of what we *do*, *think*, or *say* is often the first step toward change. The above question is also a useful tool in analyzing the solution attempts people or groups put in place.

Imagine a scenario where Bob is feeling undermined by his colleague Mike. Bob is convinced that Mike is out to get him, mistreats him, and wants to tarnish his reputation in front of their boss. What could Bob do or think to make his situation worse? For example, being aggressive with Mike or complaining about him in front of their boss. Or thinking that there is nothing to be done since Mike is already the boss's favorite and this situation can never change. Sometimes, the solution to a problem lies in reversing or stopping the actions that exacerbate it. If the boss thinks Bob is hostile toward Mike, what if Bob shows kindness and cooperation? Maybe the boss would start to question Mike's words.

Let us consider the example of an exercise called "how-to-make-it-worse," as presented in Table 2.6, which pertains to the case of Bob and Mike.

| **PROBLEM** |
| --- |
| Bob is feeling undermined by his colleague Mike. Bob is convinced that Mike is out to get him, mistreats him, and wants to tarnish his reputation in front of their boss. |
| **HOW TO MAKE IT WORSE** |
| – Bob could lose his time meticulously documenting any minor mistakes or errors made by Mike.<br>– Bob could escalate the conflict by reacting aggressively.<br>– Bob could contribute to a toxic atmosphere by engaging in passive-aggressive behavior, sarcasm, or other negative communication styles.<br>– Bob could openly criticize and belittle Mike during team meetings or in front of their boss.<br>– Bob might deliberately interfere with Mike's projects or tasks, sabotaging his efforts and causing professional setbacks.<br>– Bob could actively work to isolate Mike from team activities, conversations, and decision-making processes. |

**Table 2.6**. How to make it worse.

Paradoxes come to the rescue when dealing with sometimes overwhelming emotions, such as anger or fear. When we experience anger, it is natural to want to suppress it and keep it under control. However, this approach can often backfire, causing the anger to build up and intensify until it explodes like a cat trapped in a sack. Surprisingly, the best way to reduce anger is to give it an outlet, a *time,* and a *place*.

One effective technique is to set aside a specific time each day to focus on our anger. By writing down our thoughts and feelings, we will try to exacerbate our anger as much as possible, at that particular time and place, to prevent anger from spilling over into other areas of our lives. By allowing ourselves to feel and express our anger in a controlled way, we can learn to manage our emotions more effectively and avoid the negative consequences of suppressed anger. We will gather everything we have written without ever rereading it. When we are ready, we will throw it away.

A similar paradoxical approach can be used when dealing with fear. When we are afraid of something, such as flying, we are often told that we are safe, that there is nothing to fear, and that we should think that everything will be fine. However, this is generally ineffective because it does not consider our perceptions and subsequent reactions. Thinking that everything will be fine once again conceals fear. Instead, following the logic of paradox, we could give ourselves physical space and time to focus and think about the *worst possible scenario* in vivid detail. We would visualize everything going wrong, focusing on the most terrifying aspects of our fears and the worst possible outcomes. Repeatedly facing these fearful images can be remarkably effective in reducing their power over us and desensitizing our response. It is much like seeing the monster for the fourth time in a horror movie – the fear loses its grip, and the once-frightening image becomes almost mundane.

### 2.5.3. Unyielding beliefs

In some situations, people are so entrenched in their beliefs that any attempt to introduce change is often futile. In such cases, we must approach change with creativity, considering their beliefs and understanding them thoroughly without imposing our view of reality. It is like seeing colors differently; there is no right or wrong way to see them.

The system has to be changed *from within*, considering that a belief itself is not necessarily wrong. It becomes a problem when it hinders positive change in individuals and organizations, preventing them from growing and unlocking their talents. When someone is emotionally blocked and has a distorted perception, we must work to restructure it. Yet, we will probably not be successful by asking them to put themselves in someone else's shoes. The blocks are so strong that taking on other perspectives is often impossible. In such cases, we must work delicately, entering the other person's reality. We explore some possible techniques to achieve this.

One such technique is to *anticipate* the resistance of the other person. By acknowledging that what you are about to say may not be well-received, you can prepare your interlocutor and begin to break down their resistance,

"I will tell you something you may not like..."

At the same time, you could *lie by telling the truth*. For example, if a colleague has pulled a fast one on you, you can point it out without explicitly accusing them. By saying something like,

"If I didn't know that you loved me and would never do anything against me, I would think that you wanted to make me look bad to the boss,"

you can convey your message with less resistance.

Introducing *doubt* is another effective technique. For example, we might approach an insecure person who is convinced they are being bullied, when in fact they are not, by telling them,

"If you had not fully explained the situation to me, I might have assumed that your colleague's behavior was due to insecurity rather than a desire to harm you".

We can once again avoid triggering resistance by planting seeds of doubt in a person's mind without directly challenging their perspective.

The technique of doubt is often paired with that of *confusion*. If someone is completely entrenched in their beliefs, using complex language and non-linear logic can be a useful way to shake up their perception. Once confusion has been generated, providing simple and understandable points that support the desired change can be highly effective. It is important to note that these approaches should be used with care and consideration. It is not about manipulating or deceiving others but about encouraging critical thinking and openness to new ideas.

Strong beliefs can sometimes lead to the development of *rituals* that can become obsessive and make us feel like we will fail if we do not perform them. This is similar to an actor who feels the need to perform certain superstitious gestures before a performance. To help them, one approach is to gradually *introduce small variations* to the ritual, such as getting a word wrong or repeating the formula backward. Seeing that the performance still turns out well, the actor may begin to revise their belief. Another approach is to ask the person not to perform the ritual, while acknowledging that this may be difficult and they may fail. However, if they cannot resist the urge, they can perform the ritual with the condition of repeating it multiple times (Nardone & Portelli, 2013). By providing incentives for not performing them, individuals can gradually overcome their obsessive rituals – ultimately leading them to change their original beliefs.

At times, individuals find themselves trapped in their thoughts, unable to move forward due to the endless questions they ask themselves. Unfortunately, they lack the necessary elements to provide answers, leaving them feeling stuck. In such situations, it is the weight of thought that impedes action. Therefore, it becomes essential to *block the recurring questions* and thoughts or lead individuals to exhaust their internal dialogue. As beautifully explained in Tolle's book (2004), some of our problems stem from our tendency to obsessively dwell on the past or worry about the future, hindering our ability to embrace and experience life fully. Cultivating consciousness and presence in our daily lives and learning to accept the present moment as it is can be crucial in quieting the mind. By observing ourselves from an external perspective, detached from our thoughts, we can gradually reshape our beliefs and perceptions.

Often, fear feeds beliefs, causing individuals to struggle with decision-making or taking action. In such cases, two useful alternatives exist. The first involves *shifting the focus* away from the primary problem or blockage. For example, imagine asking a new driver to sing their favorite song while driving to work. By focusing on singing, they become distracted from the fear of driving, arriving at their destination without even noticing the journey. The second approach involves *eliminating alternatives*, similar to a commander who arrives in a foreign land to conquer it. The commander burns their ships, leaving their soldiers with no alternative but to fight and win. By removing all other options, individuals are forced to take action.

A final approach we propose relates to the concept of self-fulfilling prophecies that we have already seen. It involves acting *as if* something is attainable, even if we initially believe it is impossible. The key is not just imagining a different scenario but actively living *it out* in small steps. Just as those seeking faith benefit from attending church and praying, even before they firmly believe in God, employees can benefit from adopting a similar mindset. For example, an employee who believes they are not up to the tasks assigned by their boss may become unmotivated, unwilling, or blocked in their performance. However, by taking small steps each day, even the most trivial ones, *as if* they were an expert in their assigned work, they can build confidence and face their tasks with more courage and energy.

### 2.6. A path of change

Wanting to combine the elements presented so far, we can say that a change process should begin with clearly identifying the problem to be solved or the goal to be achieved. This initial step sets the foundation for the subsequent analysis of attempted solutions, exceptions, and the eventual agreement on the final goal and strategy to be implemented.

This strategy should not be dictated by those leading the change but rather should evolve through collaborative discussions and subsequent small agreements with the person or team involved. The desired outcome will emerge from dialogues with the individuals or teams undergoing the change. We want to emphasize the importance of asking questions and engaging in meaningful conversations rather than giving direct indications. The change path is a collaborative discovery (Nardone & Salvini, 2018).

Subsequent strategies should also be developed collaboratively, considering the resources available to those involved. If there are solutions that have worked in the past, a *guided resolution approach* may be appropriate. However, if the obstacles are more emotional in nature, such as issues related to self-awareness or interpersonal relationships, a strategy involving *corrective emotional experiences* may be more effective.

Once a strategy has been agreed upon, monitoring the situation closely to evaluate its impact is crucial. The next step involves measuring and making necessary corrections. It is important to acknowledge that the suggested actions may not always be implemented, and when they are, there is no complete guarantee of success. If we realize that a wrong turn has been taken, there is no harm in implementing a correction. Rather than viewing unsuccessful strategies as failures, they should be seen as opportunities to gain additional insights into the problem at hand.

Upon resolving the problem or achieving the desired goal, it is beneficial to reflect on the journey that led to the change. This reflection allows individuals or teams to better understand the processes involved and feel empowered by their accomplishments.

It is also essential to conduct follow-up assessments over time to ensure that the change has not resulted in a new set of problems. This is akin to avoiding trading one addiction for another, as maintaining a healthy balance is key to long-term success.

Solving a problem can significantly improve the quality of people's lives by freeing up valuable time. This newfound time must be utilized effectively to ensure continued progress. Additionally, it is crucial to reinforce any changes by consistently implementing the actions that initially led to resolving the underlying problem. By consistently practicing these beneficial actions, they will become ingrained habits, with the possibility of unlocking our untapped potential.

Figure 2.4 illustrates the process just described.

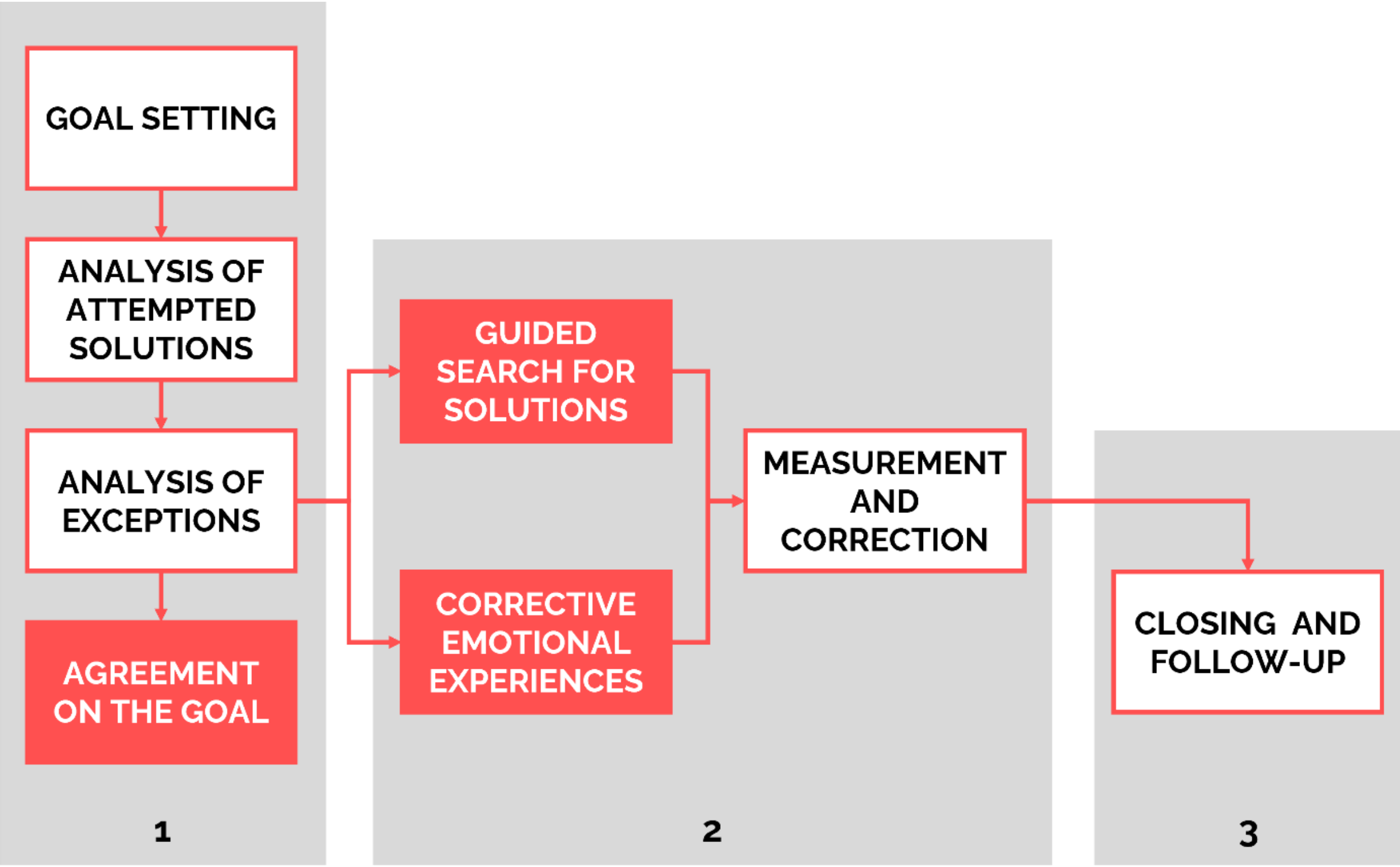

**Figure 2.4.** A path of change.

## Breaking Through Change Barriers

Prof. Andrea Fronzetti Colladon, Department of Civil, Computer Science and Aeronautical Technologies Engineering, Roma Tre University, Via Vito Volterra 62, 00146 - Rome, Italy. ORCID: 0000-0002-5348-9722

Prof. Francesca Grippa, Northeastern University, 360 Huntington Avenue, Boston, MA, 02115, USA. ORCID: 0000-0003-3537-6326

**Abstract**

This chapter explores the psychological and emotional barriers to change, highlighting emotions' critical role in overcoming resistance. It introduces different behavioral profiles encountered during change processes, from cooperative individuals to those resistant due to fear, anger, pain, or rigid beliefs. Additionally, it discusses twelve characteristics of influential change leaders, such as systemic thinking, courage, and wisdom, which are crucial for guiding teams through challenging transitions. The chapter emphasizes the need for tailored interventions, such as creating corrective emotional experiences, and offers strategic communication techniques to restructure perceptions. Through real-life examples and paradoxical approaches, it outlines practical methods to manage opposition and foster collaboration, ensuring sustainable change by addressing emotional dynamics alongside rational problem-solving.

**Keywords:** change management; change resistance; leadership; heuristics and biases; paradox of choice; self-expression.

In the previous chapter, we discussed some of the most common mechanisms that contribute to the emergence and recurrence of problems, including personal and interpersonal conflicts, psychological, social and practical issues, and everyday challenges. It should now be evident that resistance to change is not exclusively a logical process. At times, breaking through change barriers demands a profound emotional effort that surpasses our rational comprehension of the problem. Merely understanding how a problem works does not guarantee our ability to take action or find a solution. We may require new emotional experiences to catalyze meaningful change.

When analyzing resistance to change, we may encounter various types of individual or group behavior (see the profiling scheme presented in Figure 3.1). The people who appear the easiest to lead through change are those who are willing to *cooperate*. These individuals possess the emotional and practical skills necessary to unlock their talent or solve a problem, but they require guidance because the problem may be too daunting or confusing. In such cases, the first step is to establish collaboration and ensure that those collaborating do not *covertly oppose* the change. Collaboration should be tested in small doses by assigning tasks of gradual complexity to individuals or groups and *checking the results*. Once collaboration is certain, the entire path forward can be planned together, breaking the main problem into many micro-steps.

Other people *desire to change but are unable to do so*. Their limitations often stem from the emotions that drive their decisions. For example, imagine someone who wishes to jump out of a window and into the sea while their house is on fire, but the fear of heights paralyzes them. These individuals require assistance through *corrective emotional experiences*, which is much less rational than helping someone willing to cooperate. We must engage with people's emotions, initiate a dialogue to *restructure* their perceptions, and use evocative language that can "touch their hearts". By doing so, we can help them overcome their emotional barriers and embrace change.

Some individuals *intentionally* resist change. This resistance can stem from various reasons, such as *misalignment* with the assigned goal. For example, an individual may prioritize personal gain over group benefit. Alternatively, they may always take a contrarian stance in meetings to demonstrate their unique perspective to their supervisor, even if it involves dismissing good ideas presented by others. Opposition can also stem from issues related to the people driving the change effort, especially when they are not well-liked or trusted. This can occur when a supervisor has failed to earn the respect of their team or when a colleague has behaved inappropriately. At other times, resistance to change is due to a lack of *time* or *resources*. The individual may want to change, but they simply do not have the energy or time to devote to the effort required.

Opposing those who oppose is a futile exercise that only leads to conflict and escalation. Nonetheless, we must hold a firm stance, demonstrating neither weakness nor indecision, as we pursue change. Showing strength does not mean denying our limitations or presenting an illusion of invincibility. Instead, it is about embodying courage. The person who is transparent, unafraid to seek assistance, and committed to a shared goal inspires confidence, even among opponents. By embracing opposition, we can transform obstacles into learning opportunities and demonstrate our commitment to overcoming challenges on our journey forward.

The logic of paradox suggests that we prescribe resistance itself to *kill the snake with its own venom*. For example, if someone opposes every idea related to a strategic project for our company, we can *ask them to oppose* more and more, putting them in a no-win situation. We can thank them for their criticism and passionate way of pointing out the weaknesses of our project. We will explain that identifying potential problems early on is crucial for our company. By addressing critical points in advance, we can strengthen our solution and make it unassailable. We can encourage opponents to interrupt us at any time to point out any issues so that we can consider them thoroughly. Interestingly, this approach creates a paradox for anyone opposing our idea: by identifying possible flaws, they inadvertently help us reinforce the very concept they seek to challenge. Furthermore, we welcome this opposition, as it aligns perfectly with our commitment to continuous improvement.

Alternatively, we can lie while telling them the truth. We can say, for example,

“If I didn't know how much you care about the company and this project, I would think you were trying to sabotage us. But instead, I know you are only doing this to help us find a foolproof solution.”

Lastly, we can run a gauntlet that instills in the person a desire to prove themselves to us. We can say,

“Look, there is a way to solve the problem, but I don't want to propose it because it would feel like asking too much of you.”

Rigid views or unyielding beliefs can create a last type of resistance to change. In these situations, individuals can neither cooperate nor resist, as they are trapped in their inflexible mindset. To overcome this challenge, we must tap into their emotions and understand their perspective without dismissing or contradicting them. Through dialogue, we can reframe and modify their perceptions, for example, by introducing doubt or demonstrating that there are alternative viewpoints. We may even need to create initial confusion to jolt them out of their stupor and encourage them to consider new possibilities. By approaching these situations with empathy and strategic communication, we can help individuals break free from their rigidity and embrace change.

We once had a colleague – let us call her Martha – who had a peculiar way of dealing with problems. Instead of solving them, she would blow them up and put the blame on others. It was not because she wanted to hurt anyone but because she felt underestimated and ignored. Martha craved attention and wanted to prove her worth by showing how much she worked on various problems. We made the mistake of always bringing her new and effective solutions, which she promptly criticized or tried to demolish. We did not realize that by doing so, we were preventing her from proving to herself and others that she could handle the situation better. This was until we tried a different approach. We presented Martha with additional constraints that could aggravate the problem and make it difficult to solve. We then told her that solving the problem was impossible. Not to our surprise, she found an effective way out shortly after that. For her, the most important thing was to be acknowledged as the savior – the smartest, the only one capable of finding a solution.

Similarly, when a colleague or friend keeps complaining incessantly about a situation they have built for themselves and from which they do not want to or cannot get out, we might say,

“You are right. There is no escape. Now, you must decide whether continuing to complain will help or if it might be better to transform the cage you are in to make it less unpleasant.”

It is clear that in all but the first of these cases (people who cooperate), explaining *why* things happen is not the final solution. Restructuring perceptions is far more effective. Figure 3.1 presents a flowchart illustrating the categorization of a person resisting change based on their emotional and practical resources, desire for change, and whether their resistance is intentional.

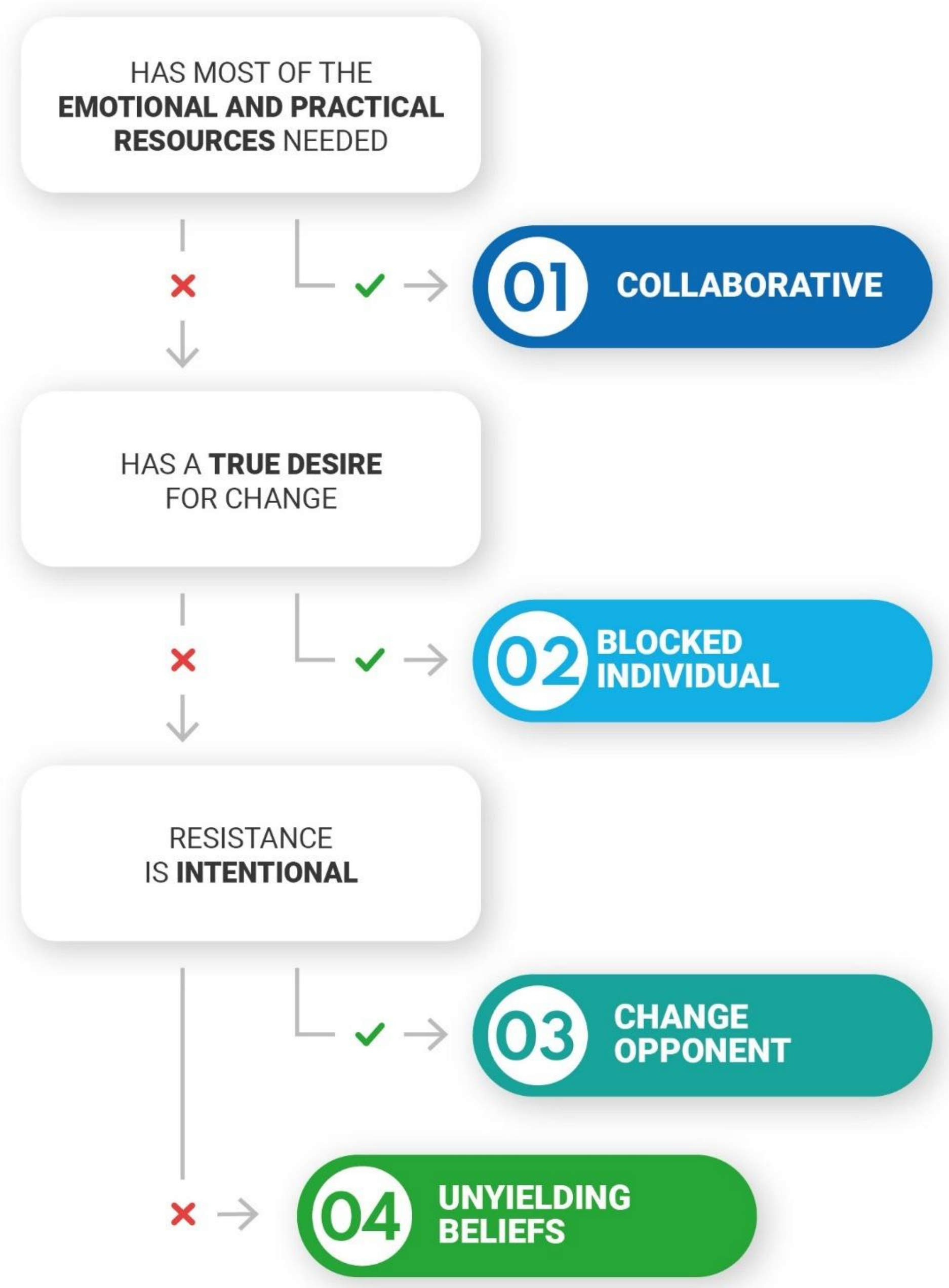

**Figure 3.1.** Profiling people who resist change.

### 2.7. Leveraging emotions

When we understand the emotions that drive individual behaviors, achieving a restructuring of perceptions and change is much easier. It is important to recognize that daunting problems often have an emotional component that cannot be resolved through purely rational means.

These emotions often stem from feelings of *anger*, *fear*, *pleasure*, or *suffering* (here intended as *emotional pain*), each of which has both positive and negative aspects that should be addressed in specific ways.

Starting with anger, it is important to note that it can have negative effects when we cannot *control* it and end up exploding in situations where we should not or where our reaction is disproportionate to the offense we have suffered. When anger is directed toward ourselves, it can also *take away our ability to react*. Anger can blind us and *dull our perceptions*, making us feel so angry that we cannot feel anything else. The most common attempted solution is to suppress the anger, which will eventually come out, often in other unrelated areas of our lives. We have already discussed how anger can be managed by letting it flow, for example, by writing down what makes us angry and trying to intensify it while confining the writing to a specific space and time. Additionally, we have learned from the logic of paradox that kindness can be a powerful weapon to dismantle those who make us angry.

Fear can most likely block us or push us to seek help and constant reassurance. When fear blocks us, we lose confidence in ourselves and do not even realize the resources we have at our disposal. Therefore, fear, like anger, can dull our perceptions. The most common attempted solution in dealing with fear is *avoiding what scares us*, trying not to do that thing, delegating it to others, or *asking for help*. In other cases, we try to *control the fear or the situation* without having all the elements to do so. For those who are afraid of something, we must block requests for help and avoidance, making them understand that every time we avoid or ask for help, we feel relieved, but we confirm our own incapacity, and therefore, the fear will only grow. By facing fear in small doses, we can reduce it. For example, for those who suffer from vertigo, we could make them climb one more step of a spiral staircase every day and then make them go back down. Only one step more each day until they reach the top. Lastly, it is helpful to avoid trying to control the situation, often by distracting the subject or shifting their attention to something else.

Consider the example of a manager who may experience fear of public speaking before an important presentation. This fear can lead to freezing during the presentation or having a trembling voice, making the presentation less effective than desired. The most common attempted solutions are obsessively controlling the presentation's content or memorizing it word-for-word. However, these methods can lead to panic and freezing at the first unexpected event or forgotten line. Some write everything they need to say on a piece of paper, resulting in an unnatural presentation and the note becoming a hindrance. Others delegate the presentation to a colleague, missing the opportunity to showcase their abilities and confirming to themselves that they are not capable. In such a scenario, we could dedicate some time in the days before the presentation to think about the *worst that could happen*, such as freezing during the presentation, saying the wrong things, or being ridiculed by the audience. The solution is to reduce our fear through paradoxical logic by giving ourselves a specific time and space to explore the worst-case scenario. Like seeing the monster in a horror movie makes it less scary, imagining the worstcase scenario ahead of time desensitizes us to our fear and makes it easier to face; in addition, it confines the fear to a specific moment in time and space, preventing it from spilling over into the rest of our day.

If our body trembles while presenting, a trick could be to hold a pen cap in our hand and squeeze it, causing a little pain. The physical pain sensation will distract us from trying to control our body, sending a signal to the brain. Finally, we could use the technique of *declaring our fear in advance*, for example, by telling the audience that we are about to discuss new topics on the frontier of research and that if there are any criticisms or different points of view, we will be willing to address them at the end of the presentation. These statements would reduce the fear of receiving criticism by establishing a more personal connection with the audience.

Dealing with suffering requires a willingness to face it head-on. Emotional pain can be debilitating, causing us to react negatively and become depressed. It can also dull our perceptions, leading us to make poor decisions based on our emotional state. This is often the case for people who, due to disappointment or failure in a past relationship, convince themselves they are better off alone, even though they might genuinely benefit from sharing a healthy relationship with a new partner. The most common reactions to emotional pain are denial or distraction, but ignoring pain only prolongs the inevitable. The only way to truly confront suffering is to go through it and learn to accept it. Whether it is the end of a relationship or the failure of our own company, we must allow ourselves time to process and move on from the pain. We can set aside time each day to reflect on what is causing us pain, use a dedicated place (like a room in our house), and try create mental images of our painful memories, making an effort to also find something positive within them. Those who help others deal with suffering can reframe the situation by asking,

“Is it better to disinfect the wound, even if it hurts, or to bleed out?”

This is a metaphor that highlights the difficult but necessary choice between short-term pain and long-term damage. It is not meant to be answered literally, but rather to make a point or provoke thought. "Disinfecting the wound" represents doing something painful or uncomfortable in the moment—such as confronting a problem, telling a hard truth, or making a tough decision. "Bleeding out" represents avoiding the issue, which may feel easier temporarily but leads to much worse consequences in the long run. It is also important to remember that we do not suffer over things without value, and if something has value, it deserves our attention, time, and care.

Ultimately, pleasure might appear as a purely positive emotion. However, it is also the most difficult emotion to work on when people find themselves trapped in its negative aspects. Extreme pleasure can be represented by addiction, such as an investor who plays and loses all their money in the stock market or a manager who loves their work and spends almost all their time in the office, neglecting their family. Pleasure can *take away our ability to react*, as we cannot stop doing what gives us pleasure or resist temptation. Those who experience pleasure often have their perceptions dulled by the *illusion that they can stop whenever they want*. The most frequent behaviors are indulging in what brings us pleasure, feeling overwhelmed by it, or attempting to control these pleasurable experiences without success. We need to be in control of pleasure instead of letting it control us. To do so, we can change the *timing and space* of pleasure. For example, someone who smokes a cigarette after lunch could force themselves to smoke one every hour, even when they do not feel like it. Or someone who smokes a lot could analyze which cigarettes are truly enjoyable during the day and focus on smoking only those, eliminating the others. Sometimes, we need a *greater pleasure* that can replace the pleasure that hurts us and creates problems.

Additional examples are outlined in Table 3.1., which presents different scenarios and associated emotions, behaviors, and potential solution strategies.

In the first row of the table, the identified problem is that the director of a department exhibits a strong need to maintain absolute control, resulting in a reluctance to delegate tasks. The prevailing emotion associated with this behavior is fear. The attempted solution involves the director refusing to delegate, insisting on doing everything himself, and constantly raising the bar for perfection. This almost leads him to burnout. The proposed strategy to address this issue involves scheduling controlled moments of planned loss of control. This could include deliberately relinquishing some tasks or responsibilities periodically. An alternative approach is to engage the director in long, unplanned meetings to create distractions, gradually prompting him to delegate tasks as he becomes occupied with other matters. In this way, he might realize, even on an emotional level, that some of the delegated tasks would be completed correctly by his colleagues, thus reducing his fear of losing control.

In the second row of the table, the identified problem is that Olivia's colleague, driven by envy, consistently opposes her during meetings in a sneaky manner and tries to undermine her reputation with the boss. The prevailing emotion associated with this situation is anger. Understanding the malicious game, Olivia responds with open attacks during meetings, which the colleague manipulatively uses to paint Olivia as an aggressive person. A possible solution strategy involves helping Olivia recognize that expressing anger in meetings demonstrates weakness rather than strength. Olivia should be encouraged to refrain from open confrontation. Instead, she could use the logic of paradox, asking her colleague to criticize her ideas and identify weaknesses, thus collaborating to make them stronger.

In the third row of the table, the identified problem is that Marco, a talented designer who loves his job, has become his boss's favorite person by taking on too many work activities that he enjoys. The prevailing emotion associated with this situation is pleasure. However, Marco soon realizes that staying late at the office every night is causing him to sacrifice too much in other areas of his life. The initial pleasure transforms into anger or pain, as he feels trapped in his obligations, and leaving early might be perceived as a drop in performance. The proposed strategy aims to guide Marco in regaining control of the enjoyable aspects of his life. One approach is to replace the pleasure derived from work with greater external pleasures, such as engaging in activities with a partner or participating in a sport he loves. Another strategy involves *ritualizing pleasure to take control of it*. For example, Marco could establish a daily routine of staying in the office until 8 p.m., even when unnecessary, and leaving not a minute earlier or later. In doing so, what was previously pleasant will no longer be so, activating a push to break current patterns.

In the fourth row of the table, the identified problem is that a manager left a previous job to start a successful startup, but serious mistakes led to the company's failure. The prevailing emotion associated with this situation is pain, which has left the manager feeling stuck at home and unable to pursue another job. The attempted solutions include moments of denial and attempts to distract oneself, but the pain persists. The proposed strategy suggests helping the manager deal with the pain by confronting it directly. This involves acknowledging the pain and guiding her through a structured process of grieving. One specific approach is to allocate half an hour each day to contemplate past mistakes, providing a defined space and time to address and process the emotions associated with the failures.

| PROBLEM/GOAL | PREVAILING EMOTION | ATTEMPTED SOLUTION AND ITS EFFECT | POSSIBLE STRATEGY |
|---|---|---|---|
| The director of a department has a very strong need to have everything under control at all times and, as a result, wants to do everything himself without ever delegating. | FEAR | He does not delegate; he does everything himself, even if he complains. He always raises the bar higher on what he demands of himself and demands that everything be perfect. This almost leads him to burnout. | Schedule, once a day, a small, controlled loss of control. Or ask his boss to keep him busy in long, unplanned meetings that can distract him so that he is gradually forced to delegate something. |
| A colleague who envies Olivia's abilities always opposes her during meetings in a sneaky way and tries to make her look bad to the boss. | ANGER | Olivia, who understands the malicious game, cannot help but blurt out in anger and openly attack her colleague during the meetings. In contrast, the colleague keeps calm and is satisfied by making Olivia appear aggressive. | Olivia could be helped to understand that every time she blurts out in anger in a meeting, she demonstrates weakness and not strength. One strategy might be asking her colleague to criticize her ideas, identify their weaknesses, and help her make them unassailable. |
| Marco is a talented designer who loves his job so much that he becomes his boss's favorite person. By taking on too many activities that he enjoys, he finds himself staying | PLEASURE | Marco initially had the illusion that he could stop working late whenever he wanted, but in reality, he could not leave early because his obligations were now too many. In addition, his | Marco must be guided to regain control of the enjoyable parts of his life. For example, by replacing the pleasure of work with a greater external pleasure, such as a partner or a |
| late at the office every night. Only to later realize that he is setting aside too much of the rest of his life. | | boss is now used to it, and if Marco left early, this would probably be perceived as a drop in performance. The initial pleasure turns into anger or pain. | sport he loves to play. Or by ritualizing pleasure in order to take control of it. For example, by forcing himself every day to stay in the office until 8 p.m., even when there is no need, and to leave not a minute earlier or later. |
| A manager left his previous job to found a successful startup but then made serious mistakes that led the company to failure. | Pain | The pain keeps her stuck at home, unable to look for another job. Sometimes, she tries to deny what happened, to tell herself it was not her fault. Other times, she tries to distract herself, but the pain comes right back even stronger. | Help to deal with the pain by going through it. For example, by making her understand that "You do not grieve for something that is not worthwhile, and if something is worthwhile, it deserves its proper place". One way might be to concentrate the pain into a half-hour a day, used in a defined space and time, to contemplate one's disasters. |

**Table 3.1.** Emotions underlying problems.

## 2.8. Good communication as a path to change

Effective communication is crucial throughout the change intervention process. The language we use, its style, and its verbal, nonverbal, and paraverbal components play a key role in conveying the change itself. We will focus on the key elements of effective communication, rather than covering every detail.

First, we recognize that rational language, which *describes* and *instructs* people on what to do, is not always sufficient for solving problems or unlocking talent. When the mechanism that fuels a problem depends on emotions, communication problems, or interpersonal dynamics, a language that can touch the heart of our interlocutor is necessary to prepare their intellect. We must enter their perception-reaction system to change it from within (Nardone, 2015). Therefore, we should use a language that can evoke emotions.

For example, to a very talented manager with a successful career who still faces opposition from rivals, we might say that "peaks, by their nature, attract lightning". We will use analogies, metaphors, aphorisms, and stories to change the mind by touching the heart. We will pay attention to everything that accompanies the content of the messages we send, such as paraverbal and non-verbal elements, like tone and rhythm of the voice and pauses to emphasize relevant parts of the speech. This book will not go into a detailed discussion of these aspects, which have been widely addressed in the literature (e.g., Nardone & Salvini, 2018; Watzlawick, Bavelas, et al., 2011). What is important to us is that language is used to create an *aversion* to solutions that do not work and *exaltation* toward new paths that can lead to change.

Engaging in meaningful dialogue requires us to do more than just assert our own opinions. Instead, we must actively seek to comprehend the perspectives of our interlocutors without imposing our own (biased) views. We must *ask much more than affirm*. We must strive to be complementary rather than imitative or confrontational. This may involve guiding those who tend to ramble with concise statements or providing broader narratives to those with overly pragmatic and narrow viewpoints. Please note that this approach is significantly different from other techniques, such as building rapport (Nickels et al., 1983; Tickle-Degnen & Rosenthal, 1990) by matching the body language of the person we are speaking with, maintaining eye contact, and matching terminology, or breathing rhythm. We aim to create attunement, not to mirror the other person's behavior. We maintain the importance of connecting with the other person rather than simply mimicking their behavior. We will see in another chapter how mirroring people's behavior is a powerful tool for change but can also lead to undesired results.

This approach involves guiding the conversation with carefully selected *questions* that will lead to a path of change. By *actively listening* to the other person, we will make them feel heard and build trust. Only then will they feel like listening to us. We will use a *soft approach* to overcome resistance and find new perspectives together. We will circumvent resistance with softness, without necessarily always having to agree with our interlocutor, but by finding new points of view together. Our goal is to *win without fighting*.

We will ask targeted questions to define the goal and investigate attempted solutions and exceptions. Nardone and Salvini (2018) suggest that these questions should not be open-ended but rather formulated as either-or questions. To understand a problem, we might, for example, ask whether it always occurs or only sometimes, whether it happens spontaneously or by choice, and whether it happens only at work or also at home. To understand attempted solutions, we may ask whether one has already tried something or not, and whether one has dealt with the problem independently or has asked for help.

At several points in the dialogue, we will use *paraphrasing* to summarize and confirm our understanding of the other person's answers. This will help us gain insights and, if needed, restructure perceptions. Indeed, some of these paraphrases will allow for adding small elements that will help the person look at the problem from new perspectives. Some examples of questions that might be used to change the perception of a problem are:

- "Are you angry or disappointed?"

- "Is it just you who has such a perception of the situation or others as well?"

- "Is rebelling against something that has already happened helpful to you or not?"

- "After reacting this way, do you feel better or worse?"

- "Would you prefer that I stay here comforting you as you share what happened once again, or would you rather we use our time differently to work on finding a solution?"

Some of these questions are also helpful in understanding the emotions behind a problem. Through these questions and paraphrases, we can create a comfortable environment where the other person feels in control. We can make the other person feel like they are leading the conversation and coming up with their *own solutions*. This approach is much more effective than simply giving advice or prescribing behaviors that may be seen as imposed from the outside. The key is to summarize the *small agreements* made along the way. Doing so can create a roadmap for change based on *joint discovery* and mutual understanding. This approach is much more likely to lead to lasting change than simply telling someone what to do. Indeed, "people are generally better persuaded by the reasons which they have themselves discovered than by those which have come into the mind of others"[7].

When engaging in dialogue, it is important to remember some general principles. Honesty is not always required, but misrepresenting ourselves or posturing will have a negative effect. For example, we might use a personal story as a metaphor for a problem even if we have never personally experienced that story. Like good actors, we will have to play the part. Often, it will be useful to add some mystery to our narrative, tying in myths and legends and remembering that a little incoherence and ambiguity can be much more fascinating than revealing everything right away. Absolute views should be avoided to prevent resistance. Therefore, we should refrain from expressing opinions, beliefs, or judgments in an overly rigid or dogmatic manner. Additionally, we should tailor the duration of our conversations to suit the individual we are engaging with, just as we would adjust the length of a lecture according to students' grade level and attention span. Finally, an evocative opening and closing can help to capture attention and make our message more memorable. By following these principles, we can create a comfortable and engaging environment that leads to meaningful change.

[7] Quote from Pensées (1671) by Blaise Pascal.

If we think about it, the crux of everything we have seen so far and of change management techniques lies in our ability to fully understand the perception of reality of our interlocutors, their consequent (re)actions, and the resulting mechanisms that fuel problems. Making wise use of communication to interrupt attempted dysfunctional solutions is the first fundamental step from which all changes start.

## 2.9. Leaders of change

When identifying the ideal individual to lead a change process within an organization, we look for someone who embodies strength, humility, and balance. This person is capable of weathering the storms that inevitably come their way. This solidity becomes the foundation upon which effective leadership is built. However, a cautionary note arises when considering leaders who lean too heavily toward charisma. Such individuals may not align with the needs of companies aiming to preserve a more collective identity rather than being overly tethered to a singular figure. Striking a balance between individual charisma and the broader identity of the organization is imperative for sustainable and adaptive leadership.

An equally crucial consideration emerges in the context of generational transitions within businesses, particularly family businesses. The assumption that the offspring of founders are inherently equipped to assume leadership roles is a potential pitfall. Too often, a life of privilege and ease can hinder their ability to navigate the challenges inherent in leadership positions. Or there may be a misalignment between skills and passions, leading to reluctant successors who lack the drive and vision necessary to propel the organization forward. Thus, a discerning approach to succession planning is vital, ensuring that the chosen leaders possess not just lineage but the requisite qualities to steer the organization through change.

A good leader is also someone who maintains a healthy balance between work and personal life, avoiding the pitfalls of being a workaholic. This balance not only enhances their well-being but also sets a positive example for their team, fostering an environment where employees feel encouraged to achieve their own work-life harmony. Work-life balance helps mitigate workplace stress by addressing various life aspects such as work, family, friends, health, and spirit/self. Leaders who apply this holistic approach, perhaps through tools like the Wheel of Life, can identify and address imbalances, thus reducing stress and improving overall productivity and satisfaction (Byrne, 2005).

In addition to balance, leaders of change exhibit eleven other essential abilities, as illustrated in Figure 3.2 and explored in the following discussion. These leaders excel in delegation, control, feedback, communication, charisma, vision, protection, motivation, courage, practical wisdom, systemic thinking, and balance.

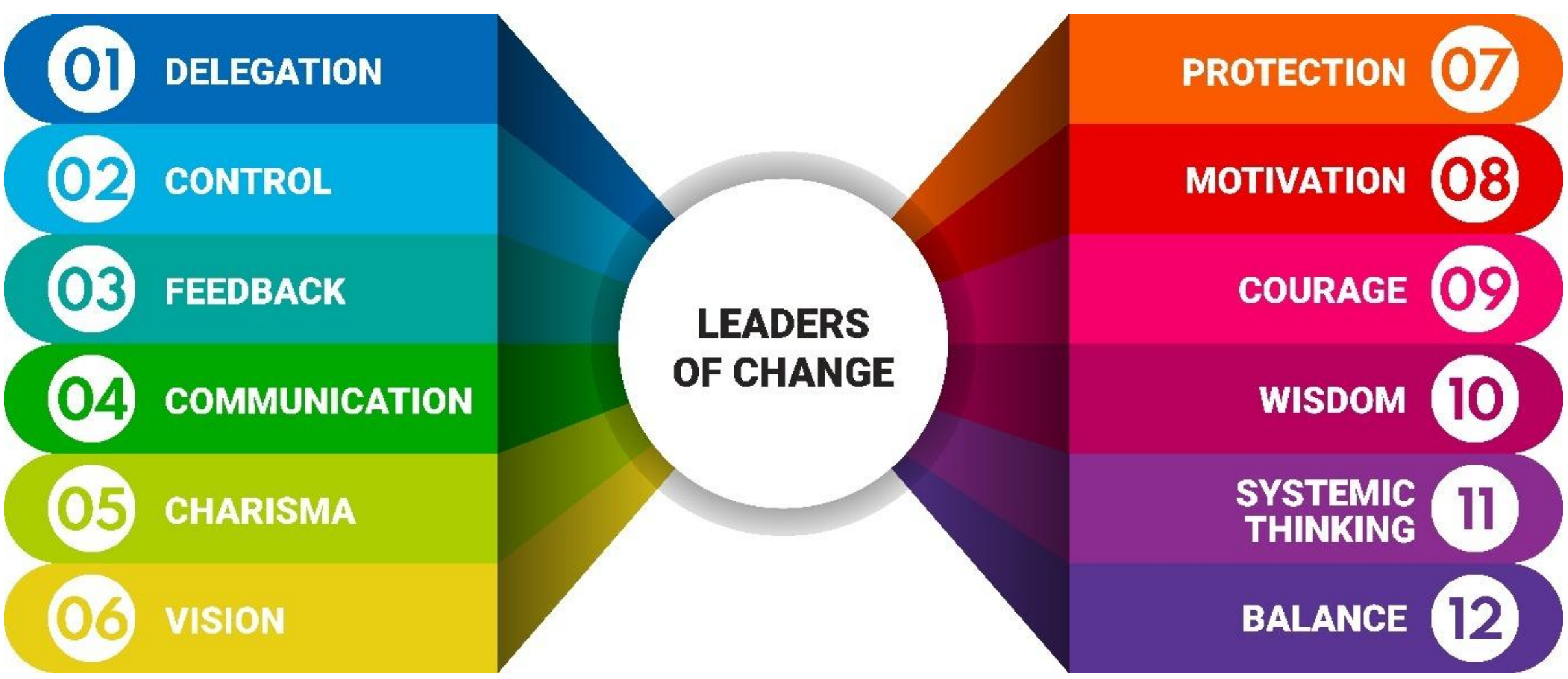

**Figure 3.2.** Leaders of change.

*Delegating responsibility.*

Delegating is a cornerstone of effective leadership, requiring a meticulous approach to defining the 'how,' 'what,' 'when,' 'where,' and 'to whom.' In this intricate dance, the leader must establish a clear boundary – delegating what remains after determining *non-delegable* responsibilities. Choosing the right individuals for delegation is an art, one that involves testing competence before entrusting tasks. The peril of becoming a centralized leader looms, a risk of shouldering every burden until inevitable burnout. The illusion of indispensability can take a leader down a treacherous path, believing they must do it all to maintain their perceived value.

Delegation, however, is not merely a distribution of tasks; it is an investment in skill development. A strategic leader understands the value of nurturing talent through delegation, fostering growth in their team. Great leaders assign challenging tasks that are within reach, enabling individuals to enhance their skills and build self-efficacy through successful completion.

Common pitfalls in delegation include being *too lenient*, which can result in tasks not being properly monitored, or being *too authoritarian*, where the focus is on punishing mistakes rather than providing guidance. Another risk is being *overly democratic*, as it may cause team members to rely too heavily on seeking assistance, ultimately hindering the potential for personal growth that effective leadership can foster. Another frequent mistake is delegating tasks we are either unable to perform or simply dislike, even when those tasks are strategic and should remain under our direct responsibility.

The key lies in striking a balance – neither hands-off nor overbearing. Effective delegation is a delicate equilibrium where responsibility is shared, growth is cultivated, and the leader retains a guiding hand without stifling the team's autonomy.

*Exercising Control With a Systemic View.*

Control, a paradoxical companion to delegation, embodies responsibility and vigilance. While empowering others, a leader must resist the temptation to believe in complete self-organization.

At the outset, control should be more palpable, and a vigilant presence should be maintained to ensure the alignment of actions with objectives. However, this intensity should gradually wane over time, evolving into a nuanced approach that tailors control to the team's specific needs. Effective control is a personalized endeavor, not a one-size-fits-all solution. It is about understanding the team dynamics, recognizing individual idiosyncrasies, and adapting the control level accordingly. The risk of "free riding," where some team members exploit the efforts of others, highlights the importance of personalized control.

A cautionary note surfaces regarding the compromises a leader may make out of fear of rejection. When this fear becomes the driving force, the leader refrains from necessary oversight, potentially compromising the effectiveness of their leadership.

Having a systemic view of the organization is also crucial for effective leadership. This means comprehending not just the individual components of the company but also how they interact and impact the system as a whole. By adopting this approach, leaders can make informed decisions that consider the long-term impact on all facets of the organization, ensuring sustainable growth and resilience. This systemic view helps anticipate outcomes of decisions and actions, aligning them with the organization's broader goals (Senge, 1990).

*Providing Constructive Feedback.*

In the art of leadership, correction emerges as a pivotal tool, complementing praise in a nuanced dance of guidance and improvement. The mantra of *praising in public and correcting in private* encapsulates this intricate balance (Blanchard & Johnson, 1983).

Feedback, inherently personal, demands a focused approach, delving into individual performance rather than painting the entire group with broad strokes. Even in the face of mistakes, the emphasis is on fostering a constructive dialogue rather than resorting to a negative or punitive tone. Correction is an opportunity to ignite change by asking for different actions without assigning blame. The leader's role is not to accuse with a blunt "you were wrong" but to engage in an analytical discussion.

"Let us analyze what we did. Did it work? If not, let us explore *how*".

Therefore, the approach to correction is akin to a thoughtful analysis, understanding both successful and unsuccessful outcomes. It involves asking questions that prompt introspection rather than assigning guilt. Acceptance of errors is vital, particularly in innovative endeavors. The leader emphasizes that innovation inherently involves risk and encourages a culture where mistakes are stepping stones to improvement. Effective leaders inspire individuals to act with courage, fostering a culture that promotes initiative and bold action. They recognize and reward those who take risks and explore new ideas, valuing effort and learning over immediate success rather than punishing failure.

While there may be instances where a stern talking-to is necessary, it is a measure to be deployed sparingly and strategically. Before contemplating dismissal, a leader must exhaust avenues of correction, ensuring that every effort is made to guide team members toward growth and improvement.

Lastly, we must consider that leaders can make errors and learn from them, demonstrating that feedback is a two-way process. Good leaders understand the importance of not only giving feedback but also receiving it. They continually evolve and adapt through learning and reflection, embracing a mindset of continuous improvement. This way, they can respond effectively to changing circumstances, ensuring that their leadership remains dynamic and capable of guiding the organization through both challenges and opportunities.

*Communicating strategically.*

Communication is the lifeblood of effective leadership. Great leaders communicate thoughtfully and actively listen to the voices of others. They balance empowerment and a healthy level of hierarchy. They maintain clear boundaries to ensure that professional relationships remain collegial rather than becoming overly personal. Strategic communication involves not only articulating a clear vision but also navigating the potential pitfalls of role reversal – a delicate overturning of established roles.

While the management system known as "Holacracy" has been celebrated as a revolutionary method as it calls for everyone to be turned into a leader, we also know employees and managers become accustomed to traditional hierarchical structures and may struggle to adapt to a completely decentralized approach. A transparent and well-communicated division of roles, with porous but clear boundaries, can help with possible accountability issues. Maintaining accountability can be challenging without clear hierarchical lines, making performance management and decision-making more difficult (Robertson, 2015).

A strategic communicator leverages the methods and tools presented in this book. They lead by example and set the tone and direction for the organization. They create an environment that welcomes diverse perspectives but understand the importance of asserting authority when needed. In this intricate dance, a leader not only provides space for others' opinions but also knows when to stop unproductive discussions.

The ability to set shared objectives is paramount, aligning the team's efforts toward common goals. A leader must effectively communicate to address any resistance within the team, especially when faced with significant changes likely to be met with opposition. A leader must navigate these challenges with finesse, avoiding going against the organizational culture, as radical changes rarely lead to successful outcomes. Instead, a leader should advocate for a more subtle approach, gradually introducing the "virus" of change to infect the organization with a positive and forward-thinking mindset.

Change leaders also demonstrate high emotional intelligence (Goleman, 2020), navigating the complexities of transition with empathy and understanding. They possess self-awareness, recognizing and managing their own emotions while attuning to the feelings of others, which fosters trust and promotes open communication. These motivated individuals leverage their emotional skills to effectively guide their teams through uncertainty, inspiring resilience and adaptability in the face of change.

*Charisma, Vision, and Motivation.*

Charisma and vision stand as twin pillars, wielding immense influence. However, it is essential to tread carefully, recognizing that the most charismatic leaders may also be prone to delusions. Their strong belief in their words can be convincing, yet it demands scrutiny. Charismatic leaders, while capable of rallying others with their compelling narratives, should be cautious not to succumb to delusion. The power of persuasion should be wielded responsibly, aligning with a clear and rational vision.

Leaders with wise charisma and vision harness intrinsic motivation, inspiring beyond rationality. Their tenacity and persistence become driving forces for the entire organization. The ability to see beyond the immediate horizon distinguishes them – a mindset that echoes the mantra: "Nothing is impossible". This should not be mistaken for the willingness to pursue a goal by disregarding ethical boundaries or overlooking the potential negative consequences of certain actions. Instead, it involves having a clear vision of what is best for the group or the organization and demonstrating the persistence and courage to achieve that vision, even in challenging times or difficult circumstances.

The use of intrinsic motivational levers sets these leaders apart, tapping into the deeper aspirations of individuals. It is not just about convincing or offering a better salary. It is about inspiring a shared sense of purpose that transcends logic or immediate financial gain. Examples of intrinsic motivation include granting autonomy, and empowering employees to take ownership of their work, which fosters independence and creativity. Motivational leaders also provide opportunities for continuous learning and skill development, sparking curiosity and personal growth. By connecting everyday tasks to a larger, meaningful mission, they help employees understand how their contributions make a real difference. Furthermore, by offering regular, genuine feedback that acknowledges accomplishments, these leaders reinforce employees' sense of personal achievement and fulfillment.

*Courage and Transparency.*

Being a courageous leader does not mean being impulsive or reckless. Just as it would be foolish to jump into a fire to impress someone on a first date, it would be equally reckless to lead a company into a high-risk situation solely to undermine a competitor, jeopardizing its future.

The ancient Greek philosopher Aristotle teaches us that true courage is not the absence of fear but the ability to face it appropriately. Courage involves confronting fear and danger because it is the right thing to do, not out of blind boldness. Excessive courage can lead to overconfidence or reckless actions without properly evaluating risks and consequences. Those who rush into danger often do so without valid reasons, driven by ignorance, vanity, or a thirst for glory rather than sound judgment.

Courage also involves humility – being unafraid to admit one's limitations and willing to learn from others. Leaders benefit from constructive feedback and should surround themselves with challengers — or better yet, sparring partners (Verganti, 2017) — who offer innovative perspectives rather than yes-men who only flatter. This openness to diverse viewpoints drives progress and helps transform outdated systems.

Courageous leaders communicate transparently, even in difficult situations or when sharing hard truths. Demonstrating humility by admitting what they do not know or asking for help strengthens their credibility, as long as the team already recognizes their competence. Similarly, courageous leaders are not afraid to apologize when they make mistakes. Instead of covering up errors, they embrace transparency, earning greater trust (Detert, 2022). We connect more with those willing to admit their flaws and tend to like them far more than those who pretend to be infallible heroes. A leader who demonstrates humility and transparency is valued for their honesty and often appears stronger, not weaker.

*Protection and Group Identity.*

Leadership extends beyond authority; it involves a commitment to protecting those entrusted under one's guidance. People naturally gravitate toward those who provide a shield in times of need, fostering a sense of security. The leader’s mantra becomes, “Touch one of mine, and you touch me”.

A true leader advocates for their team, even when team members make mistakes, handling internal issues discreetly. The key is addressing and rectifying problems from within, showing that protection does not equate to avoiding accountability but signifies a dedication to personal and collective development.

Central to this dynamic is the establishment and preservation of group identity. The leader’s protective stance becomes a powerful incentive for team members. The understanding that the group will rally to shield its members fosters a culture where individuals feel secure, motivating them to contribute confidently.

In the face of external challenges, a leader’s protective instinct not only fortifies the team but also serves as a beacon for others. Witnessing the strong support within the group, outsiders are incentivized to join, knowing that they, too, will find protection and opportunities for growth.

*Practical wisdom.*

Being a good leader does not always mean being popular or constantly pleasing the senior leadership team. It requires the courage and wisdom to prioritize what truly matters. This often involves making difficult decisions that may leave some people dissatisfied due to limited resources or differing viewpoints. Effective leaders identify flaws in outdated systems and challenge the status quo, driven by a commitment to the greater good and a clear vision of where they are headed.

A wise leader understands that an excess of rules can stifle creativity and breed a culture of rigidity. They recognize that people will inevitably find loopholes within any set of regulations. Instead of relying solely on rules, they implement incentives or *nudges* (Thaler & Sunstein, 2008) that align individual interests with collective goals.

*Practical wisdom* (Schwartz & Sharpe, 2010) gives them the courage to do the right thing and, more importantly, lets them figure out what the right thing is. They possess the moral will and skills to navigate complex situations ethically. Knowing when to bend the rules, they prioritize the greater good over strict adherence to protocol. Their leadership involves *improvisation* and exception-finding, always serving overarching objectives rather than self-interest.

By avoiding the demoralizing effects of excessive rules and incentives, good leaders cultivate a professional environment driven by intrinsic motivation rather than external rewards. They ensure that individuals are motivated to act based on their own beliefs and values rather than solely for a reward. They exercise caution in the use of incentives, recognizing that an overreliance can foster dependency, akin to a child coerced into a sport they do not enjoy simply for the promise of rewards, echoing the pattern of parental influence. Effective leaders understand the importance of fostering a supportive environment where colleagues are willing to assist each other, even when their efforts go unnoticed.

An extraordinary display of practical wisdom can be seen in the actions of Anton Schmid, an Austrian soldier who bore witness to the atrocities committed against Jews by the Nazi regime. Despite the risks involved, Schmid made the courageous decision to defy orders and provide Jewish families with food, shelter, and false papers to help them escape persecution. His selfless acts saved the lives of hundreds of Jews before he was eventually caught and executed in 1942. Schmid's courage and compassion serve as a shining example of humanity during the darkness of the Holocaust.

A study by Gloor et al. (2022) explored the potential of artificial intelligence in analyzing the communication of individuals, particularly employees, to gauge their ethical and moral attitudes. Drawing a parallel with the animal kingdom, the authors differentiated between "ethical" behaviors, likened to bees, related to individuals who care about a greater good, and "moral" behaviors, likened to ants, as a metaphor for individuals who tend to follow group or community rules without question. Additionally, they identified leeches as selfish individuals solely focused on their own interests. A leader who embodies practical wisdom is akin to a bee, prioritizing the collective good over personal gain. Research indicates that such leaders play a crucial role in enhancing group performance. While individuals who adhere strictly to rules without questioning them, like ants, may receive higher ratings, having leaders who prioritize the greater good is essential for overall success (Gloor et al., 2022).

We continue this chapter with some additional sections that highlight research essential for understanding human behavior. Although these sections are concise relative to the extensive studies they summarize, they enhance and complement the knowledge we have gained thus far.

## 2.10. Heuristics and biases

In this section, we touch on fundamental theories that shed light on specific aspects of human behavior, specifically heuristics and biases, and what motivates individuals to change. Specifically, we refer to the works of Kahneman (2013), Thaler and Sunstein (2008), and Schwartz (2004a), which are of enormous significance in this field. We aim to contextualize these studies within the discourse presented thus far and explore how biases and heuristics can sometimes be used to support the change management process.

As Kahneman (2013) illustrated, our brain functions with two distinct systems. System 1, operates quickly and intuitively, making decisions with minimal effort. It can be thought of as our automatic pilot, responsible for most of our daily actions and choices. System 1 relies on heuristics and intuition to make rapid judgments and decisions. It continuously scans our environment for potential threats and opportunities, linking ideas and concepts based on similarity and proximity. This system helps us form associations between different stimuli and recognize patterns in our surroundings. On the other hand, System 2 is slower, more logical, and requires conscious effort. It is often lazy and indecisive, but it is essential for complex tasks that require problem-solving and critical thinking.

System 1 allows us to react quickly to immediate dangers, such as running, because the tree we are under is about to fall. System 2 helps us plan and execute more complex actions, such as choosing the best escape route in the event of a falling tree. Thanks to System 1, we can perceive distance and hear sounds, while System 2 allows us to measure distance and distinguish between different sounds.

Although System 1 and System 2 are both essential for decision-making, they have their respective advantages and disadvantages. While System 2 requires more effort, energy, and time to make a decision, System 1 is prone to errors and influenced by our habits and recurring behaviors that might fuel problems. It relies on norms and expectations to make quick judgments and jump to conclusions based on limited information. Stereotypes and preconceptions often influence it. On the other hand, System 2 can actively lead us to reflect and change our perspective, considering multiple factors and causes of a problem. It is more open to surprises and new information, making it a valuable tool for decision-making. In practice, the two systems often work together, with System 1 providing an initial impression or intuition and System 2 providing a more detailed analysis and evaluation of the information.

When guiding someone through a change process, it is important to be aware of their reliance on mental shortcuts or rules of thumb, known as heuristics. If someone is too focused on their automatic responses, we must help them break out of their patterns and become aware of their heuristics. For example, reacting angrily is a typical effect of a decision made with System 1, where anger takes control.

Heuristics can often lead to biased decisions. For example, we may draw conclusions based on a small sample of data and assume that all carrots are orange simply because we have only seen orange carrots. Our brains also tend to rely too heavily on the first piece of information we encounter when making a decision (Tversky & Kahneman, 1974). This initial information is known as the *anchor* and is frequently used in negotiations. In negotiations, the anchor refers to the first price or offer made by a party, and it serves as a subconscious reference point for both parties throughout the discussion. The anchor can have a powerful effect on the outcome of the negotiation. For example, a merchant who wants to sell a chandelier for at least $30 will be much more likely to succeed by starting the negotiation at $100 rather than $35.

When faced with a complex question, our brains may substitute it with an easier one (Kahneman & Frederick, 2002). A professor who is asked to judge if a student has a high GPA only from their picture might substitute the question with “does this person look like a good student?” Similarly, we tend to perceive events as more likely to occur when they are linked to strong emotions like fear or anger. We also tend to judge the likelihood of an event based on how easily we can recall examples of it. If we can easily think of examples of a particular event, we are more likely to judge it as more likely to occur (Tversky & Kahneman, 1974).

There are numerous heuristics and biases that we could discuss at length. Cialdini (2006) described six famous principles frequently used in marketing and sales. Interestingly, these same principles can also be used to guide change.

One principle is *reciprocity*. It suggests that people feel a sense of obligation to repay others for what they have received. By giving first, we can create a sense of indebtedness that can be used to influence others to reciprocate. For example, in a coaching intervention, we can establish trust by being the first to share our personal stories, creating a welcoming environment that encourages the other person to open up and share their experiences in return.

Another principle is related to commitment and *consistency*. People often desire to be consistent in their beliefs and actions. By asking them to take a small action, we can create a sense of commitment that can be used to ask for larger commitments in the future. This is why we usually plan a change and personal growth process starting from the smallest possible steps. Once a person embarks on this path and makes some initial progress, they are more likely to put in additional effort to reach their goal.

A third principle is *scarcity*. People tend to value things that are rare or in limited supply. By emphasizing the limited availability of a particular product or opportunity, we can create a sense of urgency that can be used to influence people's behavior. This principle is often used in commercial offers with deadlines or when booking a hotel room online, where the website informs us that it is the last room available, and we must hurry. In a change intervention, we could stress the importance of the person's chance to work on themselves, as their company has decided to invest in their training and growth – a time-limited opportunity offered to only a select group of employees.

The three additional principles are *social proof*, *authority*, and *liking*. Social proof is the concept that people are influenced by the actions of others. By demonstrating that a particular behavior is popular or endorsed by others, we are more likely to follow suit. For example, we may choose a crowded restaurant over an empty one, even if the latter is cheaper and better, but may be empty as it has only been open for two days.

Liking refers to the idea that we are more likely to say yes to people we like. This is not limited to physical attractiveness but also includes the tendency to like people we feel are similar to us, for example, because they have gone through similar experiences or because they share our values and beliefs.

Lastly, authority suggests that people are more likely to follow the lead of those who are perceived as experts or authorities. To effectively guide individual or group change, we must establish ourselves as respected and knowledgeable in our field. This can be demonstrated, for example, through publications or a proven track record of success.

According to these last three principles, we would be more effective in leading individual or group change if we were known as respected and experienced in our field (authority), if we could show that many other people have received our help (social proof), and if we could adapt to different contexts by getting in tune with our audience (liking).

## 2.11. When choosing is a problem

It is often mistakenly believed that having many choices is a good thing. However, having more options often means having to make more decisions, which can be difficult or scary. The difficulty in making decisions often stems from the desire to find the perfect choice that involves no risk or error. Unfortunately, this attitude can lead to decision paralysis, inaction, and ultimately, giving up on making any choice at all. The problem is that decisions always have consequences, which can create anxiety when faced with a crossroads. This problem can be exacerbated when faced with important decisions, such as buying a house or choosing a college major.

In today's world, we are constantly bombarded with choices, from what to wear to what to eat and which streaming service to subscribe to. Schwartz (2004a) illustrates that having too many options can lead to feelings of stress, anxiety, and dissatisfaction. Even if we are generally satisfied with our decision, we may feel that we have missed out on something better. This is often referred to as the fear of missing out on other options, which is aligned with the view of a *liquid society* where some individuals tend to avoid choices or relationships that are perceived as permanent (Bauman, 2003).

Reducing the anxiety behind a decision can sometimes be achieved by realizing that it does not have to be definitive and that many decisions are reversible. However, revising our decisions still has a cost that individuals often do not want to face. One approach to limit our options is to set rules or eliminate certain options altogether. Another useful method is to imagine future scenarios linked to our decisions to evaluate how we would feel if a decision were made. This can help us better understand the consequences of our choice and evaluate whether it is the right one for us.

In an effort to support change, we must also work to reshape individuals' perceptions and help them become aware of their fears and the mechanisms that fuel them. Often, when we struggle to make decisions, it is because we fear failure or *losing control*. To overcome this, we should focus on the positive aspects of our choices and be grateful for what we have rather than worrying about what we may miss out on. This is easier for people whom Schwartz (2004a) classifies as *satisficers,* who are satisfied with good enough options, rather than individuals who strive for the best possible outcome. *Maximizers*, on the other hand, try to make the best possible decision by carefully evaluating all available options, but this requires a lot of time and effort and, paradoxically, often leads to anxiety or regret about one's decision. Also because of loss aversion, "when they compare themselves with others, they get little pleasure from finding out that they did better and substantial dissatisfaction from finding out that they did worse" (Schwartz, 2004b, p. 72).

In general, it is advisable to stop our search when we find some good options, even if they may not be the absolute best. It is important to manage our expectations and refrain from dwelling on the options we did not choose. Furthermore, it is crucial to recognize that it might be beneficial to restrict our options when the choice is not crucial. For example, when selecting a perfume, we could limit ourselves to sampling no more than ten fragrances from the wide array available at the perfume store.

To promote societal well-being, policymakers should present choices in a way that encourages better decision-making without forcing it. People, including employees, should be provided with an appropriate level of choice, which varies depending on the specific field and their level of engagement. For example, when onboarding new employees and selecting office supplies, a computer scientist should be provided with more options than another employee who simply needs a computer for basic tasks like writing letters. Supporting individuals in their decision-making involves providing a carefully curated set of options from the vast array available, guiding them through the complexity of excessive choices. Furthermore, these options should be organized thoughtfully to encourage wise decisions.

Thaler and Sunstein's (2008) concepts of *libertarian paternalism* and *nudges* suggest that *choice architects* have the possibility (and responsibility) to simplify the decision-making process, presenting choices in a way that makes it easier for people to make good decisions. This can be done by providing clear and easy-to-understand information, simplifying the decision-making process, and making the default option the most beneficial for the individual. The default option is especially important for those who avoid making decisions.

For example, setting the default option to paperless billing can encourage individuals to reduce paper waste and improve the sustainability of their actions. Similarly, many appliances and devices could come with energy-efficient default settings, thereby promoting eco-friendliness.

Moreover, nudges can support people even when default options are unavailable. Nudges refer to small changes in how choices are presented or framed that can influence people's behavior toward a certain direction without restricting their freedom of choice (Thaler & Sunstein, 2008). For example, placing cigarettes in a part of the shop that is hard to find for customers and healthy foods right at their fingertips can be considered nudges.

Understanding the mechanisms of change and how our perception-reaction systems function is the first step in facilitating wise decision-making. At the group level, leaders should steer their teams toward optimal choices by offering well-organized and relevant options. Finally, in the era of big data, artificial intelligence can also serve as a coach, learning our habits and recommending tailored choices that best fit each individual's needs.

## 2.12. Understanding and leveraging the power of self-expression

As discussed in this chapter, understanding individual and team perceptions and reactions is crucial to driving meaningful change. In order to set goals and identify the correct problems to be solved, we must first comprehend the motivations of individuals and the (repeated) mechanisms that fuel problems. It is important to remember that we constantly communicate with others, ourselves, and the world around us. Throughout this continuous communication process, social norms and culture significantly influence our behavior. Therefore, it is also essential to investigate the characteristics of the environment and society in which individuals are embedded.

Among the numerous works conducted in this field, we choose to briefly touch on the sociology of consumption. The concept is that the objects we choose to buy and use can reflect our personal tastes, preferences, and values, and, more importantly, that they can serve as a means of expressing our individuality and identity to others (Ferraro et al., 2011; Roberti, 2021; Warde, 2017). As such, objects become integrated into our language. To illustrate this concept, we had some fun and asked ChatGPT[8], "How should I dress to go to a concert at a community center if I wanted to fit in well with the context?". Part of the answer was:

*"In general, it is a good idea to dress neatly and conservatively for a community center concert. This might mean wearing clean, pressed clothes in subdued colors, such as a pair of nice slacks*

[8] https://openai.com/blog/chatgpt. Accessed on March 2023.

*or a skirt with a blouse or button-down shirt. [...] While you may want to express your personal style and individuality through your clothing, it is generally best to avoid wearing anything too flashy or distracting to a community center concert. This might mean avoiding large logos or patterns, excessive jewelry, or anything that might draw attention away from the music or the performers".*

We subsequently asked if it was ok to wear a designer bag, and the answer was:

*"[...] While it is possible to wear a designer bag or other luxury accessory to this type of event, it may be seen as excessive or out of place".*

Similarly, we asked, *"How should I dress to join an Emo community?"*. Part of the answer was:

*"To dress in an Emo style, stick to a dark color palette, mainly black, with accents like deep reds or purples. Wear skinny or ripped jeans, band t-shirts, and layer with studded belts, plaid shirts, or hoodies. Converse or combat boots complete the look. Hair is typically straight, with long, side-swept bangs, often dyed black with bold streaks of color. Heavy eyeliner and accessories like chokers, wristbands, and piercings add to the aesthetic. Emo fashion is about expressing emotion and individuality".*

We then asked whether it was acceptable to wear fluorescent colors and floral patterns, and part of the response was:

*"Fluorescent colors and floral patterns aren't typically part of the Emo aesthetic, which leans heavily toward dark, muted tones and edgy designs. Emo fashion is about expressing emotion and individuality, but it generally avoids bright or flashy colors".*

Our clothing choices are not solely based on comfort or functionality but also serve to express our personal style. However, it goes beyond just style, as our clothing also communicates our values. For example, we may wear clothing showcasing our commitment to protecting the environment or reflecting our political ideology. Additionally, our clothing can signify our membership in a particular social group, as we often dress similarly to those within our circle to demonstrate our shared values and sense of belonging. It is important to note that social and cultural norms heavily influence our preferences and choices. As such, what *we consider* fashionable or desirable is often shaped by the values and beliefs of those around us.

Self-expression is a key component in consumer choices. It highlights a sociological perspective often overlooked during a new product's design and development stages. Purchasing decisions are no longer just a response to a specific need but have become a social fact. This is particularly evident when analyzing the behavior of teenagers today compared to those of previous decades. Objects once considered superfluous are now *essential* for leading a *normal* existence and being accepted, recognized, and respected by peers.

What matters now "is the moment of acquisition", not the lasting friendship with the object. Possessions are no longer signs of permanence and stability; they are often consumables, things we use and discard. Objects go out of fashion as quickly as they appear on the market. In the liquid modern world, it is the style that needs to be kept alive at all times, not its accessories (Bauman, 2010). Accordingly, brands can effectively guide purchase choices by enabling consumers to represent a piece of their identity.

The transition from the modern to the post-modern era has emptied the public space of normative and cultural references that once structured people's identities. In the past, work was considered a primary characteristic of individual identity. However, in today's world, our identities are shaped across more complex dimensions. We are no longer defined solely by our jobs but by a variety of interests and beliefs. For example, someone may be a scientist, a Buddhist, a scuba diving enthusiast, a vegan, and a Harley Davidson rider all at once. This shift in social conditions has led to a more fluid society, which sociologist Zygmunt Bauman (2000) calls "liquid modernity". The rigid containers of collective identity are dissolving, making way for more flexible and adaptable individual realities. This constant redefinition of self requires a significant amount of effort.

Dahrendorf (1979) maintained that the opportunities of choice available to individuals ("options") gain meaning through "ligatures" – which are strong bonds that link individuals to society. Ligatures allow us to order and select our options by setting criteria of relevance, discarding those that are empty and meaningless. Our individual perspectives take shape within the framework of values in which they are embedded. In preindustrial societies, people had strong social linkages but few options. In contrast, these ligatures have weakened in today's liquid world, sometimes leading to a substantial indifference between choice options, as the individual *can be anything*. This has resulted in a deficit of values, leading to a more flexible but less stable sense of self. A more fluid and less restrictive society requires significant effort to redefine ourselves constantly. Everyone now strives to construct their own identity, including through consumption choices. "The Western type of individualized society tells us to seek biographical solutions to systemic contradictions" (Beck & Beck-Gernsheim, 2002, p. xxii). As individuals, we have the freedom to shape our identity, but it comes at a cost. We become responsible for our identity, and if we are not satisfied with what we have become one day, it is harder to blame someone other than ourselves, such as society or family.

The act of consumption has taken on a significant role in defining who we are, as the products we purchase communicate our individuality to both ourselves and others. Brands have become a crucial part of this language of things, as they bring with them qualities and attributes that we incorporate into our identities. We have been persuaded that our lives are considered unsuccessful unless we stay abreast of the latest trends (Lawson, 2009). What we buy shows who we are. We purchase objects not only for their practical use but also to convey the image of who we aspire to be and how we want others to perceive us. However, consumers today are seeking more than just trendy products. They want the brands they choose to align with their values and beliefs. They desire a dialogue with manufacturing companies where their feedback and needs are taken into high consideration.

In this context, analyzing citizens' and consumers' spontaneous expressions through big data analytics presents unprecedented opportunities to study new trends and make informed business or policymaking decisions. By understanding the language of things and the values consumers hold dear, companies can create products that resonate with their target audience and build a loyal customer base.

Importantly, these considerations extend beyond personal transformation and offer valuable insights into the mechanisms of societal change. In today's world, identities are fluid, and traditional collective norms are becoming less influential. Consumer choices represent a dialogue with society, which can actively contribute to reshaping social narratives. By leveraging this, businesses and policymakers can promote positive change on a broad societal scale.

After reading the initial chapters of this book, it should be evident that communication is multifaceted and extends beyond verbal or written exchanges. Additionally, not all communication is consciously experienced or transmitted. In the upcoming chapters, we explore how to identify honest signals in communication and collaboration patterns and discuss how to intervene effectively to improve performance and foster personal and team growth. We then present advanced text mining and social network analysis tools that can aid in analyzing the behaviors of individuals, groups, and businesses.

## References

Bach, K. (1981). An Analysis of Self-Deception. *Philosophy and Phenomenological Research*, *41*(3), 351. https://doi.org/10.2307/2107457

Bateson, G. (1972). *Steps to an Ecology of Mind*. Chandler Publishing Company.

Bauman, Z. (2000). *Liquid Modernity*. Polity Press.

Bauman, Z. (2003). *Liquid Love: On the Frailty of Human Bonds*. Polity Press.

Bauman, Z. (2010). *44 Letters From the Liquid Modern World*. Polity Press.

Beck, U., & Beck-Gernsheim, E. (2002). *Individualization: Institutionalized Individualism and its Social and Political Consequences*. SAGE Publications.

Blanchard, K., & Johnson, S. (1983). The One-Minute Manager. *Cornell Hotel and Restaurant Administration Quarterly*, *23*(4), 39–41. https://doi.org/10.1177/001088048302300409

Buzan, T. (2010). *The Mind Map Book: Unlock Your Creativity, Boost Your Memory, Change Your Life*. Pearson Education Limited.

Byrne, U. (2005). Wheel of Life: Effective steps for stress management. *Business Information Review*, *22*(2), 123–130. https://doi.org/10.1177/0266382105054770

Cialdini, R. B. (2006). *Influence—The Psychology of Persuasion*. Harper Business.

Dahrendorf, R. (1979). *Life Chances: Approaches to Social and Political Theory*. Weidenfeld and Nicolson.

De Bono, E. (1985). *Six Thinking Hats: An Essential Approach to Business Management*. Little, Brown, & Company.

Detert, J. R. (2022, January 7). *What Courageous Leaders Do Differently*. Harvard Business Review. https://hbr.org/2022/01/what-courageous-leaders-do-differently

Ferraro, R., Escalas, J. E., & Bettman, J. R. (2011). Our possessions, our selves: Domains of self-worth and the possession–self link. *Journal of Consumer Psychology*, *21*(2), 169–177. https://doi.org/10.1016/j.jcps.2010.08.007

Freud, A. (1992). *The Ego and the Mechanisms of Defence*. Routledge.

Freud, S. (1989). *New introductory lectures on psychoanalysis*. Norton & Company.

Gloor, P., Fronzetti Colladon, A., & Grippa, F. (2022). Measuring ethical behavior with AI and natural language processing to assess business success. *Scientific Reports, 12*(1), 10228. https://doi.org/10.1038/s41598-022-14101-4

Goleman, D. (2020). *Emotional Intelligence: Why It Can Matter More Than IQ*. Bloomsbury Publishing.

Kahneman, D. (2013). *Thinking, Fast and Slow*. Farrar Straus & Giroux.

Kahneman, D., & Frederick, S. (2002). Representativeness Revisited: Attribute Substitution in Intuitive Judgment. In T. Gilovich, D. Griffin, & D. Kahneman (Eds.), *Heuristics and Biases: The Psychology of Intuitive Judgment* (pp. 49–81). Cambridge University Press. https://doi.org/10.1017/CBO9780511808098.004

Lawson, N. (2009). *All consuming*. Penguin Books.

Lee, M.-K. (2005). *Epistemology after Protagoras: Responses to Relativism in Plato, Aristotle, and Democritus Epistemology after Protagoras: Responses to Relativism in Plato, Aristotle, and Democritus*. Oxford University Press. https://doi.org/10.1093/0199262225.001.0001

Merton, R. K. (1948). The Self-Fulfilling Prophecy. *The Antioch Review*, *8*(2), 193. https://doi.org/10.2307/4609267

Milanese, R., & Mordazzi, P. (2014). *Coaching Strategico*. Ponte alle Grazie.

Nardone, G. (2009). *Problem solving strategico. L'arte di trovare soluzioni a problemi irrisolvibili*. Ponte alle Grazie.

Nardone, G. (2015). *La nobile arte della persuasione*. Ponte alle Grazie.

Nardone, G. (2016, December 22). *Teoria e pragmatica del cambiamento*. Psicologia Contemporanea. https://www.psicologiacontemporanea.it/blog/teoria-e-pragmatica-del-cambiamento/

Nardone, G., & Balbi, E. (2017). *Solcare il mare all'insaputa del cielo: Lezioni sul cambiamento terapeutico e le logiche non ordinarie*. TEA.

Nardone, G., & Portelli, C. (2013). *Ossessioni compulsioni manie. Capirle e sconfiggerle in tempi brevi*. Ponte alle Grazie.

Nardone, G., & Salvini, A. (2018). *Il dialogo strategico. Comunicare persuadendo: Tecniche evolute per il cambiamento*. Ponte alle Grazie.

Nickels, W. G., Everett, R. F., & Klein, R. (1983). Rapport Building for Salespeople: A Neuro-Linguistic Approach. *Journal of Personal Selling & Sales Management*, *3*(2), 1–7.

Roberti, G. (2021). Youth Consumption, Agency and Signs of Girlhood: Rethinking Young Italian Females' Lifestyles. In A. M. Vogel & L. Arnell (Eds.), *Living like a girl. Agency, Social Vulnerability and Welfare Measures in Europe and Beyond* (pp. 45–64). Berghahn Books.

Robertson, B. J. (2015). *Holacracy: The New Management System for a Rapidly Changing World*. Henry Holt.

Schwartz, B. (2004a). *The Paradox of Choice – Why More Is Less*. Harper Perennial.

Schwartz, B. (2004b). The tyranny of choice. *Scientific American*, *290*(4), 70–75.

Schwartz, B., & Sharpe, K. (2010). *Practical Wisdom: The Right Way to Do the Right Thing*. Penguin.

Senge, P. M. (1990). *The fifth discipline: The art and practice of the learning organization*. Doubleday Currency.

Thaler, R., & Sunstein, C. R. (2008). *Nudge: Improving Decisions About Health, Wealth, and Happiness*. Yale University Press.

Tickle-Degnen, L., & Rosenthal, R. (1990). The Nature of Rapport and Its Nonverbal Correlates. *Psychological Inquiry*, *1*(4), 285–293. https://doi.org/10.1207/s15327965pli0104_1

Tolle, E. (2004). *The Power of Now: A Guide to Spiritual Enlightenment*. New World Library.

Tversky, A., & Kahneman, D. (1974). Judgment under Uncertainty: Heuristics and Biases. *Science*, *185*(4157), 1124–1131. https://doi.org/10.1126/science.185.4157.1124

Verganti, R. (2017). *Overcrowded: Designing Meaningful Products in a World Awash with Ideas*. The MIT Press.

Warde, A. (2017). *Consumption: A Sociological Analysis*. Palgrave Macmillan UK. https://doi.org/10.1057/978-1-137-55682-0

Watzlawick, P., Bavelas, J. B., & Jackson, D. D. (2011). *Pragmatics of Human Communication: A Study of Interactional Patterns, Pathologies and Paradoxes*. W. W. Norton & Company.

Watzlawick, P., Weakland, J. H., & Fisch, R. (2011). *Change: Principles of Problem Formation and Problem Resolution*. Norton & Company.